\documentclass{pinchcrnmodified}   % Comment the above and uncomment this
\usepackage{graphics}   
\usepackage{graphicx}   
\usepackage{amsmath}   
\usepackage{bm}   
\usepackage{epsfig}   

\title{Coherent exciton transport in semiconductors}

\author{ Massimo Rontani }
\affiliation{CNR-NANO Research Center on nanoStructures and bioSystems
at Surfaces (S3), \\ 
Via Campi 213a, 41125 Modena, Italy}
\author{L. J. Sham}
\affiliation{Department of Physics, 
University of California San Diego, \\
Gilman Drive 9500, La Jolla, CA 92093-0319}
\authors{2}

\begin{document}

\maketitle

\acknowledgements

We thank Leonid Butov for critically reading the manuscript.
This work is supported by EU-FP7 Marie Curie Initial Training Network 
``Indirect Excitons: Fundamental Physics and Applications (INDEX)''.

\tableofcontents

\maintext

%&&&&&&&&&&&&&&&&&&&&&&&&&&&&&&&&&
\chapter{Coherent exciton transport in semiconductors}
%&&&&&&&&&&&&&&&&&&&&&&&&&&&&&&&&&

%&&&&&&&&&&&&&&&&&&&&&&&&&&&&&&&&&
\section{Introduction}
%&&&&&&&&&&&&&&&&&&&&&&&&&&&&&&&&&

An exciton is a particle-like neutral excitation of solids and molecules
composed of one electron and one hole bound together by the 
mutual electrical attraction 
\cite{Knox1963,Rashba1982,Schafer2002,Shah1999,Yu2004}. Its creation
through internal charge separation is most frequently caused by the
absorption of light and its demise is occasioned by 
electron-hole recombination, mostly with emission of light and less frequently 
non-radiatively.
The many-electron ground state of the system, being an insulator, 
is immune to excitation until the excitation energy reaches a 
threshold $G$ known as the energy gap. 
When external influences such as the electromagnetic
field and lattice vibrations are ignored, the exciton may be viewed as a  
robust state of an excited electron plus the hole which has been left
behind in the valence electron states \cite{Sham1966}. 
The hole acquires its positive charge
from the loss of an electronic charge from the ground state
whose total charge is neutralized
by that of the ions in the molecule or solid.

The photon-exciton interaction is responsible for the
optical excitation (though not necessarily in the visible frequency 
range) of the exciton and for its spontaneous recombination 
emitting a photon (a 
quantum unit of light). The dipole matrix
element responsible for the transition between the energy states
is strong when the electron and hole wave functions overlap
in space or match in wave vector. From Planck's law,  
the frequency of the emitting light $E_X/h$ is proportional to
the energy loss $E_X$
in returning the exciton state back to the ground state, with 
$h$ being Planck's constant.

If the constituent electron and hole of the exciton are mostly 
localized at an ion, the exciton is localized, but with some
probability to hop from site to site. Such a Frenkel exciton
is common in molecules and molecular solids.
At the other extreme, if the electron and hole wave functions
are widespread as extended orbitals in a molecule or Bloch
waves in a crystal, their bound state as the exciton can have 
their center of mass moving through the system with ease.
Such Wannier (or Wannier-Mott) excitons are most common
in broad-band and small-gap semiconductors (a semiconductor is 
distinguished from an insulator qualitatively by a smaller 
energy gap, with the frequency of the emitting light from the
exciton spanning the range from visible light to very far infrared).
Wannier excitons resemble the hydrogen atom or, more closely, the
positronium system composed of an electron and a positron. 
Because of the dielectric screening of the electrical force
in small-gap materials and sometimes the small effective mass
of the electron, the Wannier exciton radius is 10 to 100
times larger than the positronium radius, which is approximately 0.1 nm.

Excitons, being made of two fermions, behave 
as bosons on the scale larger than the exciton radius and therefore
may macroscopically occupy a single 
quantum state \cite{Blatt1962,Moskalenko1962,Casella1963,Keldysh1967,Agranovich1967,Keldysh1968,Gergel1968}. 
If the exciton lifetime is long enough to allow for 
reaching quasi-equilibrium, 
the dilute and cold gas of optically generated excitons 
may undergo Bose-Einstein condensation 
(BEC) \cite{Noziers1982,Griffin1995,Stringari2003}.
The critical temperature for exciton BEC, of the order
of 1 K for typical densities in semiconductors, 
is basically the temperature at which the thermal
De Broglie wavelength becomes comparable to the average 
inter-exciton separation.
The possibility of achieving BEC of excitons by shining light 
on solids has been thoroughly investigated in the last fifty years
(see the reviews \cite{Hanamura1977,Mysyrowicz1980,Griffin1995,Littlewood1996,Moskalenko2000,Larionov2001,SnokeRev2002,Butov2003,Butov2004,Butov2007,Ivanov2008,Snoke2011}). 
Semiconductors are particularly appealing for this goal
as they may provide excitons with a a lifetime (hundreds
of ns in bilayer structures \cite{Butov2003}) 
longer than the time required for cooling. 

In an indirect-gap semiconductor such as silicon, where the
momentum of the exciton does not match that of the photon, the excitons
are generally formed after relaxation of optical excitations with
initial energy much higher than the gap. The indirect exciton has a long
lifetime because its recombination with the emission of a photon 
requires the conservation of momentum to be satisfied by the assistance
of a lattice vibration or trapping by a defect. Consequently, the excitons
have time to form a large pool known as an electron-hole 
drop (see the reviews \cite{Rice1977,Keldysh1986}).
Alternatively, the delay in optical recombination may be due to the symmetry
of the crystal,
as in the direct-gap oxide Cu$_2$O which has 
conduction and valence bands of like parity  
hence the optical dipolar transition is forbidden \cite{Hulin1980}.
Chapter 8 by Kuwata-Gonokami in Volume 1 focuses on the aspects of
Bose-Einstein condensation of optically generated excitons in semiconductors.

In a direct-gap material, the spatial separation of the electron and hole
can be enforced by housing them in two layers sufficiently close to
maintain their electric 
attraction \cite{Shevchenko1976,Lozovik1976,Fukuzawa1990}. 
The recombination of such indirect
excitons may then be controlled by changing the electron and hole
wave function overlap with an
electric field \cite{Alexandrou1990,Zrenner1992}. 
An interesting phenomenon is the laser spot excitation
of these indirect excitons, leading to 
the formation of 
two concentric luminous circles centered at the laser spot plus other
localized bright spots randomly placed between the circles.
Whereas the formation of the inner ring 
\cite{Butov2002b} is due to 
the migration of indirect excitons away from the
laser spot as optically inactive excitons, the
localized bright spots \cite{Butov2002b} as well as the outer ring  
\cite{Butov2002b,Snoke2002,Butov2004b,Rapaport2004} form
on the boundaries between electron-rich and hole-rich regions. 
At low temperature
the outer ring is a necklace of evenly spaced bright spots,
whose origin is not fully understood \cite{High2012}.
This system of excitons in a double quantum well is considered a good
candidate for condensation.
These experiments and related work in 
double quantum wells are reviewed in
\cite{SnokeRev2002,Butov2003,Butov2004,Butov2007,Timofeev2007,Ivanov2008}.

Another possible---and elusive---mechanism of condensation of 
excitons as bosons is that excitons form spontaneously at thermodynamic 
equilibrium even in the absence of an optical excitation.
Such process signals the transition to a permanent phase known as excitonic 
insulator (EI) \cite{Mott1961,Knox1963,Keldysh1964,Cloizeaux1965,Kohn1967},
which is originated by the instability of the normal ground state 
of either a semiconductor or a semimetal
against the spontaneous formation of bound electron-hole pairs.
The wave function of the strongly correlated EI ground state 
is formally similar to that proposed
by  Bardeen, Cooper, and Schrieffer for superconductors \cite{BCS1957}.
As a matter of fact, both excitons and Cooper pairs are absent 
except as fluctuations in
the normal high-temperature phase and form only  
in the ordered, low-temperature phase---respectively 
the EI and the superconductor.
Besides, both condensation of excitons and that of Cooper pairs are best
described in the reciprocal space of the crystal solid. 
The EI phase is reviewed in \cite{Halperin1967,Kohn1968,Hanamura1977,Noziers1982,Noziers1982b,Griffin1995,Littlewood1996,Moskalenko2000,SnokeRev2002,Littlewood2004,Ivanov2008,Snoke2011}.

It is intriguing to observe that
condensation of other types of bosons composed of two fermions leads
to spectacular manifestations 
of quantum mechanical coherence,
such as the superfluidity ensuing from the pairing
of $^3$He atoms \cite{Legget2006}, Fermi alkali atoms
confined in optical traps \cite{Bloch2008,Giorgini2008}, 
nucleons in neutron stars
\cite{Ginzburg1964,Pines1969},
the superconductivity induced  
by Cooper pairs in metals \cite{deGennes1999}, and
the non classical momenta of inertia
in nuclei \cite{Migdal1960,Bohr2008}. The above phenomena may regarded 
as distinct realizations of superfluidity, associated
to the coherent, dissipationless flow of charge and / or mass.
However, excitons are neutral and
stay dark unless recombine radiatively, as shown in Table \ref{t:features}, 
which compares
the distinct features of the condensates made of composite bosons. 
The signature of the macroscopic order of the exciton condensate
is, at present, controversial for the superfluid transport but its
other manifestations will be discussed next. 

\begin{table}[tb]
\tableparts
{
\caption{Excitonic insulator (EI) versus superconductor \emph{{\`a} la}
Bardeen-Cooper-Schrieffer (BCS).
The interface referred to in the Table
is the junction between normal and condensed phase.
For a general discussion of the condensates made of composite
bosons see \cite{Kohn1970}. For specific EI features see
\cite{Kohn1967} (Meissner effect),
\cite{Zittartz1968a} (superconductivity), \cite{Zittartz1968b}
(superthermal conductivity), \cite{Rontani2005a,Rontani2005b} 
(Andreev reflection),
and \cite{Rontani2009} (Josephson oscillations).
}
\label{t:features}
}
{
\begin{tabular}{lll}
\hline
 & & \\[-8pt]
Physical property & Excitonic insulator & BCS-like superconductor \\[2pt]
\hline
 & & \\[-6pt]
Nature of the composite boson & Exciton & Cooper pair \\[3pt] 
Boson charge & Neutral & 2$e$ \\[3pt]
Boson momentum   
& Crystal momentum & Crystal momentum \\[3pt]
&                  & (commonly ignored in the free \\[3pt]
&                  & electron gas approximation) \\[3pt]
Boson mass & Effective mass 
                   & Effective mass of the electron \\[3pt] 
&                  & quasiparticle in the Fermi level \\[3pt] 
&                  & region (of thickness provided by \\[3pt] 
&                  & phonon Debye frequency) \\[3pt] 
Type of long-range order & Diagonal  
& Off-diagonal \\[3pt]
Superfluidity & ? & Superconductivity \\[3pt]
Meissner effect & No & Yes \\[3pt]
Superthermal conductivity & No & No \\[3pt]
Nature of the quasiparticle & Electron (hole) & Bogoliubon \\[3pt]
Andreev reflection & Yes & Yes \\[3pt]
Interface electric conductance & Decreased & Increased \\[3pt]
Interface thermal conductance & Decreased & Decreased \\[3pt]
Proximity effect & Yes & Yes \\[3pt]
Josephson oscillations & Yes & Yes \\[3pt]
 & & \\[-6pt]
\hline
\end{tabular}
}
\end{table}

The aim of this Chapter is to illustrate
some recent theoretical
proposals concerning the detection of coherent exciton flow
\cite{Rontani2005a,Rontani2005b,Rontani2009}.   
The reader may refer to the literature reviewed in Sec.~\ref{s:review} 
for a discussion of the conceptual and experimental 
difficulties inherent in
the realization of exciton condensates.   
Here we set aside such difficulties and
adopt in a pedagogical way the simplest mean-field
description of the condensate, on which we lay our theoretical
development in order to detect the transport properties of
the exciton condensate. 

In particular, we focus on the exciton analogues 
of two phenomena, i.e., Andreev reflection and Josephson effect, 
which are hallmarks of superconducting behavior, and stress the crucial
differences between excitons and Cooper pairs.   
Our first main conclusion is that the excitonic insulator is the {\em perfect}
insulator in terms of both charge and heat transport, with
an unusually high resistance at the interface with a 
semimetal---the normal phase of the condensed state.
Such behavior, which should be contrasted 
with the high electrical conductance of the junction between superconductor 
and normal metal,  may be explained in terms of the coherence induced into the
semimetal by the proximity of the exciton condensate.
Then we show that the exciton superflow may be directly probed in the case that
excitons are optically
pumped in a double-layer semiconductor heterostructure:
we propose a correlated photon counting experiment
for coupled electrostatic exciton traps which is a variation of
Young's double-slit experiment. 

We last mention that, due to the interaction between electrons and light,
not only can an exciton decay irreversibily into a photon or vice versa, but it
can also exchange roles with the photon in a quantum-mechanically coherent
fashion. Thus, the exciton may exist in the solid 
in the superposition state of an exciton and a photon,
known as polariton. Whereas the photon energy varies linearly with 
its momentum at the speed of light in the vacuum, the exciton energy  
depends on the square of its center-of-mass momentum.
For small momenta, the exciton and the photon can approximately 
match both their momentum and energy values, the coupling mixing 
the two states into
two superpositions of photon and exciton with an energy splitting. Thus,
the massless photon is slowed down by the massive exciton by virtue of the 
quantum-mechanical superposition. The Chapter by Yamamoto in Volume 1 deals
with aspects of polariton condensation.

The structure of this Chapter is the following:
After a review of previous work
(Sec.~\ref{s:review}),
in Sec.~\ref{s:twoband} we illustrate the
mean-field theory of the EI emphasizing its relation with  
the BCS theory of superconductors. 
We then introduce the phenomenon of Andreev reflection
in Sec.~\ref{s:Andreev} and analyze its observable
consequences in Sec.~\ref{s:results}.
Section \ref{s:Josephson} on the Josephson effect ends the Chapter.

\section{Physical systems}\label{s:review}

This section briefly reviews recent theoretical 
and experimental works on exciton condensation,
focusing on diverse physical systems.
Without attempting an exhaustive review,
we refer the reader to more comprehensive essays whenever available. 

\subsection{Bose-Einstein condensation of 
optically generated excitons}

The pursuit of Bose-Einstein condensation of optically generated excitons
in semiconductors, which dates back to the sixties, presently focuses on 
both classic systems such as Cu$_2$O and novel low dimensional structures
(for reviews see \cite{Hanamura1977,Mysyrowicz1980,Griffin1995,Littlewood1996,Moskalenko2000,Larionov2001,SnokeRev2002,Kavoulakis2003,Butov2003,Butov2004,Littlewood2004,Butov2007,Timofeev2007,Ivanov2008,Snoke2011,Kuwata2011,Schwartz2012}).
A very active field concerns ``indirect'' excitons. 
Such excitons are made of
spatially separated electrons and holes, hosted in two quantum wells that 
are sufficiently close to maintain electrical attraction between the carriers 
of opposite charge. This setup has several advantages: 
(i) The overlap of electron and hole wave functions is controlled by
applying an electric field along the growth direction of the 
bilayer heterostructure, thus increasing the exciton recombination time 
by orders of magnitude with respect to the single-well value
\cite{Alexandrou1990,Zrenner1992}. 
(ii) The confinement effect along the growth direction
increases the exciton-phonon scattering rate, 
improving exciton thermalization \cite{Zhao1997}. 
(iii) The dipolar repulsion among indirect excitons
disfavors the formation of biexcitons and electron-hole
droplets \cite{Butov1994,Littlewood2004,Stern2008,Schindler2008,Vogele2009,Lee2009,Voros2009,Cohen2011} as well as 
effectively screens the in-plane disorder potential 
\cite{Ivanov2002,Savona2007,High2009a,High2009b,Remeika2009,Alloing2011}.
(iv) As the electric field parallel to the growth
direction may be laterally varied 
using suitably located electrodes, 
one may tailor the in-plane effective potentials for 
excitons, thus realizing artificially controlled 
traps \cite{Huber1998,Hammack2006,Chen2006,Gorbunov2006,Gartner2007,High2009a,High2009b,Schinner2011,High2012b,Alloing2011,Alloing2012a},
ramps \cite{Hagn1995,Gartner2006}, 
lattices \cite{Zimmermann1997,Zimmermann1998,Remeika2009,Remeika2012}, 
``exciton circuits'' \cite{High2007,High2008,Grosso2009},
and ``exciton conveyers'' \cite{Winbow2011}. 

Exciton traps may also be created by means of the 
uncontrolled in-plane disorder of the double quantum well 
\cite{Zrenner1994,Butov2002a,Butov2002b,High2009a,High2009b}, 
the strain experienced
by the heterostructure \cite{Trauernicht1983,Kash1988,Negoita1999,Naka2005,Voros2006,Yoshioka2011}, 
the laser-induced confinement \cite{Hammack2006bis,Alloing2012b},
the magnetic field \cite{Christianen1998}.
The realization and control of exciton traps 
is a key capability to reach exciton BEC: As 
the long range order in two dimensions is smeared by
quantum fluctuations, a weaker requirement for the macroscopic
occupation of the lowest exciton level
is that the exciton coherence length exceeds the trap 
size \cite{Butov2004}.

The present evidence of exciton BEC is based on distinct 
features of the emitted light (photoluminescence, PL) 
that appear at low temperature:
(i) The PL dynamics exhibits bosonic stimulation
of the scattering of hot optically dark excitons into 
optically active low-energy states \cite{Butov2001}. 
(ii) The PL signal becomes noisy in a broad
range of frequencies, as it occurs in the presence of coherence 
\cite{Butov1994,Krivolapchuk1999,Kuzmin2010}.
(iii) The degree of polarization of the emitted light increases
with decreasing temperature
\cite{Larionov2000,Larionov2002,High2012}, consistently with 
gauge symmetry breaking.
(iv) The exciton mobility is enhanced, which may be attributed 
to superfluid behavior \cite{Butov1998}.
(v) The radiative decay rate increases, which may be 
explained in terms of ``superradiance'' of a macroscopic dipole   
\cite{Butov1998} or collective behavior at the onset of condensation
\cite{Larionov2000,Larionov2002}.
(vi) The PL lineshape narrows 
and departs from the Maxwell-Boltzmann distribution
\cite{Krivolapchuk1999,Larionov2000,Larionov2002,Larionov2002b},
as it may be expected for the macroscopic population of a single 
exciton state. 

However, some of the signatures 
\cite{Hulin1980,Snoke1987,Lin1993,Fukuzawa1990b}
listed above, taken separately, 
may have different explanations than 
exciton condensation 
\cite{OHara1999,OHara2000,Kash1991}
(for a discussion see Ref.~\cite{Butov2004}).
The most compelling evidence of BEC
is probably the direct measure of coherence
through interferometric techniques \cite{Yang2006,Timofeev2007,High2012,High2012b,Semkat2012,High2012c,Alloing2012c}, which accesses the macroscopic
exciton wave function in real space.
On the theory side, the light emitted by excitons just after the
onset of condensation is predicted to be
coherent \cite{Oestreich1996,Rossier1998,Castro2001},
with a sharply focused peak of radiation in the direction normal to the
quantum-well plane \cite{Keeling2004,Zimmermann2005}.
Besides, the instability leading to the
external ring of evenly placed bright spots discussed in the introduction
\cite{Butov2002b,Butov2004b} is possibly linked 
to exciton quantum degeneracy \cite{Levitov2005}.

An intriguing issue is the role played  
by spin \cite{Maialle1993} in 
exciton condensation. In bilayers, the exciton spin 
is the component $J_z$ of the  
angular momentum along the direction perpendicular to the planes, 
discriminating between optically active ($J_z=\pm \hbar$)
and inactive states ($J_z=\pm 2\hbar$). The most urgent questions concern the
multi-component nature of the condensate \cite{Rossier1997,Yao2008}, 
the possibility of dark-exciton condensation  
\cite{Combescot2007,Combescot2011}, 
the role of spin-orbit coupling \cite{Hakioglu2007,Can2009,Wu2008,Shim2009}.
This research is fueled by the recent experimental evidence that
the spin-relaxation time of indirect exciton is long and consequently  
exciton spin transport is long-ranged 
\cite{Leonard2009,Kowalik2010}, as well as that spin textures and 
polarization vortices appear together with the onset
of long-range coherence \cite{High2012}. 

\subsection{Excitonic insulator in mixed-valence semiconductors}

In principle, 
any intrinsic semiconductor that may be turned 
into a semimetal, either by applying stress or by suitable alloying,
may undergo a transition to the permanent EI phase.
Favorable conditions are the presence of an indirect gap, which 
weakens the detrimental effect of dielectric screening
on the exciton binding, as well as the nesting of electron and 
hole Fermi surfaces, which maximizes electron-hole pairing.  
Nevertheless, early experiments focusing on simple materials, such as 
divalent fcc metals (Ca, Sr, Yb) and group V semimentals (As, Sb, Bi)
were unable to confirm the existence of the EI. 
References \cite{Halperin1967,Kohn1968,Hanamura1977,Noziers1982,Noziers1982b,Griffin1995,Littlewood1996,Moskalenko2000,Littlewood2004,Ivanov2008} review 
the work on the EI.

Recently, a few experiments have pointed to the realization of the EI phase
in mixed-valent semiconductors. The first class of 
candidate materials consists in rare-earth
chalcogenides, such as TmSe$_x$Te$_{1-x}$ \cite{Bucher1991,Wachter1995},
Sm$_{1-x}$La$_x$S \cite{Wachter1995,Wachter1995b}, Sm$_{1-x}$Tm$_x$S,
YbO and YbS \cite{Wachter1995b}. These 
compounds all crystallize in the NaCl structure and
undergo a semiconductor-semimetal transition 
as the band gap $G$ is changed
from positive to negative values by applying
high hydrostatic pressure to the sample.

When the direct gap of TmSe$_{0.45}$Te$_{0.55}$,
formed between the localized 4$f^{13}$ levels and the 5$d$
conduction-band states, is closing with
external pressure, an indirect band gap
develops between the highest valence Tm 4$f^{13}$ level $\Gamma_{15}$ at
the $\Gamma$ point and the mimimum of the $\Delta_{2^{\prime}}$
conduction band 5$d$ states at the X point of the Brillouin zone.
As the otherwise localized 4$f$ band is broadened and shows
a maximum at $\Gamma$ due to $p$(Se,Te)-$f$(Tm) covalent
hybridization \cite{Jansen1985}, it is tempting to use a simple 
two-band model for
interpretation, similar to the one ilustrated in Sec.~\ref{s:twoband}.
On the basis of low-temperature
resistivity and Hall mobility measurements, the authors of \cite{Bucher1991}
attribute  the 
resistivity increase with the vanishing gap to a condensation of free carriers
into excitons, 
placing 
the EI phase between semimetal and semiconductor,
close to $G\approx 0$. 
Later, the same group has reported a linear
increase of thermal conductivity and diffusivity with decreasing
temperature and attributed it to exciton superfluidity
\cite{Wachter2004}.
Fehske and coworkers \cite{Bronold2006,Bronold2007,Zenker2012} 
have suggested theoretically that the EI phase
in the pressure-temperature phase diagram is narrower that the
experimental claim, being surrounded
by a ``halo'' regions made of preformed excitons coexisting 
with the normal semiconductor phase.
The presence of this halo, precursor of the EI, explains the
experimental findings and rules out the idea of a 
heat supercurrent, which conflicts with the general argument
\cite{Kohn1967,Zittartz1968b}
that a flowing condensate carries no entropy and thus no heat. 

Other candidate systems for the EI phase are the transition-metal 
chalcogenides TiSe$_2$ 
\cite{DiSalvo1976,Wilson1977,Traum1978,Margaritondo1981,Stoffel1982,Anderson1985,Pillo2000,Kidd2002,Cercellier2007,Rasch2008,Monney2010,vanWezel2010,Monney2011,May2011,Cazzaniga2012}, 
Ta$_2$NiSe$_5$ \cite{Wakisaka2009}, TaSe$_2$ \cite{Taraphder2011},
and the possibly ferromagnetic EI GdI$_2$ 
\cite{Bascones2002,Taraphder2008}
(see Ref.~\cite{Rossnagel2011} for a review).
The main evidence relies on the hole quasiparticle band structure,
as extracted from angular-resolved photoemission 
\cite{Traum1978,Margaritondo1981,Stoffel1982,Anderson1985,Pillo2000,Kidd2002,Cercellier2007,Rasch2008,Monney2010,May2011,Wakisaka2009}. 
The much studied TiSe$_2$,
at a critical temperature of around 200 K, develops a charge density wave 
\cite{DiSalvo1976} 
which does not fit the standard model based on Fermi surface nesting
\cite{Gruner2000,Rossnagel2011}
but it is consistent with the presence of an
EI (see Sec.~\ref{s:twoband}).
In fact, the spanning
wave vector of the charge density wave
is the distance in reciprocal space between Ti 3$d$-electrons 
and Se 4$p$-holes, which are bound by Coulomb attraction.
Therefore, the excitonic instability drives the charge density wave
and may possibly couple with a periodic lattice distortion 
\cite{Stoffel1982,Pillo2000,Cercellier2007,Monney2010,Monney2011,May2011,Cazzaniga2012}, 
though alternate scenarios 
\cite{Kidd2002,vanWezel2010,Rossnagel2011}  
have been suggested.
Recent time-resolved photoemission data link the artificially
induced collapse of the 
charge-ordered TiSe$_2$ state to screening due to transient 
generation of free charge carriers, supporting the excitonic 
origin of the phase transition \cite{Rohwer2011,Hellmann2012}.

A third class of candidate systems consists in
Kondo insulators \cite{Aeppli1992} and heavy-fermion materials
\cite{Lee1986}, 
which are mixed-valence semiconductors
characterized by a flat $f$-type valence band plus 
a dispersive---say $d$-type---conduction band,
typically exhibiting strongly correlated behavior. 
Such systems (e.g. SmB$_6$) are often
modeled by the Falicov-Kimball Hamiltonian, which takes into account
the strong inter-band Coulomb interaction 
\cite{Falicov1969,Portengen1996b,Rontani2000,Rontani2009b}.
Sham and coworkers have shown 
\cite{Portengen1996a,Portengen1996b}
that the exciton condensate made of  
$f$ holes and $d$ electrons 
may spontaneously break the lattice inversion symmetry and 
lead to a ferroelectric phase transition of electronic origin, 
whereas conventional ferroelectricity is associated to lattice 
distortion \cite{Kittel1986}.  
The predicted experimental signatures, supported
by some evidence \cite{Wachter1985,Glushkov2006}, 
include the divergence of the
static dielectric constant, a ferroelectric resonance in the
microwave absorption spectrum, and a nonvanishing susceptibility for
second-harmonic generation. 

If intraband hybridization 
dominates over Coulomb interaction, then the exciton condensate 
wave function acquires a different type of symmetry---$p$-wave---which 
excludes the ferroelectric scenario but allows the coupling
with the lattice \cite{Duan1997}. In this latter 
case the excitonic instability manifests itself
as a spontaneous lattice deformation which may explain some of the phase 
transitions known as ferroelastic \cite{Bulou1992}.

\subsection{Permanent exciton condensation in bilayers}\label{s:bilayers}

In order to investigate permanent
exciton condensation in semiconductor bilayers,
one strategy is to host electrons in the first layer 
and holes in the second layer \cite{Shevchenko1976,Lozovik1976}. 
This task is nowadays accomplished by means of
suitable electric gates which allow to separately contact the layers 
\cite{Sivan1992,Kane1994,Pholt2002,Keogh2005,Prunilla2008,Takashina2009}.
The spacer beetween the two quantum wells suppresses the inter-layer 
tunneling which induces exciton recombination, but 
it is sufficiently thin to provide strong inter-layer 
Coulomb interaction (see Ref.~\cite{DasGupta2011} for a recent review).
This setup allows for measuring the Coulomb drag resistance, which 
is the inverse ratio of the electric current measured in one layer  
to the open-circuit voltage developed in the other layer in turn.
Such drag resistance is predicted to diverge in the
presence of exciton condensation,
as the exciton binding 
correlates the motion of carriers in the two layers
\cite{Vignale1996,Hu2000,Joglekar2005}.
Recent measurements
\cite{Croxall2008,Seamons2009,Croxall2009}
point to low-temperature
anomalies in the Coulomb drag  
which may originate from an excitonic instability, though other
strongly correlated phases are possible \cite{DasGupta2011}.

An alternate strategy is to place 
electrons in both layers in
the presence of the magnetic field 
(see \cite{Eisenstein2003,Eisenstein2004,Tutuc2007} 
for reviews). The field bends classical electron trajectories
into circular cyclotron orbits. As such orbits may be placed
all across the plane, overall their quantized energies consist in highly
degenerate ``Landau levels''. Since the level degeneracy 
is the number of quanta of magnetic flux that cross the plane,
for sufficiently high fields and identical layers the lowest 
Landau level in each layer will be half filled by electrons
(single layer filling factor $\nu = 1/2$,
total filling factor $\nu_{\text{T}}=1$). Note that in this 
quantum Hall effect regime, 
routinely detected through the quantization 
of the Hall resistance,  
Landau levels may be considered either half filled
or half empty.
Therefore, one may switch to the excitonic 
parlance \cite{Kuramoto1978,Yoshioka1990},
regarding one layer as filled by electrons and the other one by holes:
In this picture the exciton
``vacuum'' has the lowest Landau level totally filled in one layer 
($\nu=1$) and empty in the other layer ($\nu=0$), 
thus excitons are created by moving electrons from one layer, 
which leaves a hole behind, to the other one \cite{Rezayi1990}. 

There is significant evidence, based on low-temperature
transport experiments, that the bilayer ground
state is a condensate of excitons. 
The first hint is a huge enhancement of inter-layer
tunneling solely due to many-body effects, clearly pointing
to strong inter-layer coherence \cite{Spielman2000}.
The most compelling observations are based on counterflow measurements
\cite{Shevchenko1976,Lozovik1976,Su2008}, where 
the electric currents of opposite sign and like magnitude
that flow in the two layers provide zero total electric current
and a net exciton flow.
For filling factors other than $\nu_{\text{T}}=1$ the 
Hall voltages separately measured in
the two layers are equal and opposite in sign, 
whereas for $\nu_{\text{T}}=1$ they both drop to zero, consistently with the
flow of an uncharged object such an exciton 
\cite{Kellog2004,Tutuc2004,Wiersma2004}.
To prevent edge states---always present at the
boundary of quantum Hall systems and unrelated to excitons---from
playing a role in transport, the 
Coulomb drag has been recently measured in the ``Corbino'' annular geometry, 
confirming the excitonic nature of transport
\cite{Tiemann2008,Tiemann2008b,Finck2011,Nandi2012},
whereas the superfluid character of the exciton flow is unclear.
The above scenario is supported by
the measurement of quasiparticle and collective excitations 
by means of tunneling \cite{Spielman2001,Giudici2010}
and inelastic light scattering 
\cite{Luin2003,Karmakar2009} spectroscopies.

Interesting theoretical predictions concern the response of the 
bilayer exciton condensate to external electromagnetic fields
\cite{Balatsky2004,Joglekar2005,Sonin2009,Eastham2012} 
and impurities \cite{Dubi2010}, as well as the transport
properties of hybrid circuits including exciton condensates and
superconductors \cite{Dolcini2010,Peotta2011,Soller2012}.
For weak inter-layer interaction 
or filling factors other than $\nu_{\text{T}}=1$, 
bilayers are predicted to
undergo phase transitions to other strongly correlated phases,
such as paired two-dimensional Laughlin liquids 
and Wigner solids \cite{Yoshioka1990},
or peculiar excitonic charge density waves \cite{Chen1991}.

\subsection{Graphene-based systems}

Graphene---a recently discovered 
allotrope of $sp^2$ bonded carbon---is a one-atom thick 
two-dimensional honeycomb lattice 
\cite{Geim2007,CastroNeto2009,DasSarma2011,Kotov2012}. 
Its peculiar electrical and mechanical properties---chemical stability, 
high mobility, easiness of making electric contacts---have stimulated 
observations by means
of different electron spectroscopies and scanning probes. 
Intensive investigations have uncovered new
physics (e.g. Klein tunnelling, 
anomalous types of quantum Hall effect), rooted in the unusual 
character of quasiparticle excitations, that, in the
neighborhood of the Fermi energy, 
are massless chiral Dirac fermions. 
In fact, conduction and valence bands
form specular cones whose apexes touch in the two
inequivalent points K and K$^{\prime}$, located at the corners of the
hexagonal two-dimensional Brillouin zone.
These two points, which map into each
other by a rotation of $2\pi/6$ \cite{Grosso2000}, are the 
Fermi surface of the undoped system, hence graphene is a zero-overlap
semimetal. 

In principle, graphene is a good candidate system for EI, since:
(i) the density of states vanishes at the charge neutrality point, 
hence the long-range Coulomb interaction is unscreened
(ii) the perfect electron-hole symmetry of Dirac cones
favors the nesting of electron and hole isoenergetic surfaces.
Khveshchenko \cite{Khveshchenko2001} was the first to
suggest that graphene hides
a latent excitonic insulator instability.
The EI phase is a charge density wave alternating between
the two inequivalent triangular sublattices, its spamming wave
vector connecting K and K$^{\prime}$ in reciprocal space. A stack
of graphite layers in a staggered (ABAB...) configuration,
with the atoms located in the centers and corners of the hexagons
in two adjacent layers, respectively, could stabilize the EI 
by enforcing interlayer Coulomb interaction.

After this seminal prediction, many theoretical works have
tried to estimate the size of the EI transport gap 
as well as the stability of the EI phase
(see for example \cite{Vafek2008,Drut2009,Gamayun2009,Gonzalez2012,Wang2012} 
and references therein as well as the reviews 
\cite{CastroNeto2009b,Kotov2012,Semenoff2012}). 
The absence of consensus is not surprising,
as the many-body problem in graphene is presently
an open issue \cite{Kotov2012,Uchoa2012}.
Experiments show that electrons in graphene allegedly 
behave as non-interacting particles 
\cite{Geim2007,CastroNeto2009,DasSarma2011}, except
for small effects related to velocity renormalization \cite{Elias2011}, 
coupling with phonons \cite{Geim2007,CastroNeto2009,DasSarma2011} 
/ plasmons \cite{Bostwick2010} 
(here we are not concerned with the fractional quantum Hall
effect \cite{Du2009,Bolotin2009}, induced by the magnetic field). 
Therefore, if the EI energy gap ever exists, it must be smaller
than the present spectroscopic resolution. 

A related theoretical proposal concerns permanent exciton condensation
in double-layer graphene \cite{Lozovik2008,Dillenschneider2008}. 
The idea is to separately contact the two layers,
which are spaced by a dielectric medium, in order to induce the
same quantity of charge
with opposite sign in the two layers. With respect to the double-layer
made of usual semiconductors mentioned in Sec.~\ref{s:bilayers}, 
here the advantage is the smaller value of the transverse
electric field required to polarize the bilayer, due to the zero
energy gap of graphene.
The estimate of the Kosterlitz-Thouless temperature
required to undergo the EI phase is debated theoretically
\cite{Min2008,Zhang2008,Kharitonov2008,Bistritzer2008,Gilbert2009,Kharitonov2010,Mink2011,Lozovik2012,Sodemann2012,Suprunenko2012,Phan2012}; 
recent Coulomb drag measurements \cite{Kim2011,Gorbachev2012}
point to the importance of inter-layer interactions.

In the absence of a dielectric spacer, undoped bilayer graphene
is predicted---among other proposals---to undergo an 
excitonic ferroelectric phase that spontaneously breaks 
which-layer symmetry and polarizes the layers in charge 
\cite{Min2008b,Nandkishore2010,Zhang2010}.
The excitonic instability, which opens an energy gap,
appears to sensitively depend on the
interaction range \cite{Guinea2010}: in the case of finite range, 
the expected electronic phase 
is nematic and gapless \cite{Lemonik2010,Vafek2010} (see 
reviews \cite{MacDonald2012,Nandkishore2012}). The experimental
observation of a transport
gap at the charge neutrality point is controversial 
\cite{Weitz2010,Mayorov2011,Velasco2012,Freitag2012}.

Other interesting graphene-based systems are carbon nanotubes, 
which may be thought of as wrapped sheets of graphene \cite{Saito1998}. 
Since nanotube electronic states are built from those of graphene 
after imposing suitable boundary conditions, they exhibit perfect 
electron-hole symmetry \cite{JarilloHerrero2005} and hence provide 
the optimal nesting of energy bands to achieve permanent exciton 
condensation. Besides, Coulomb interactions are especially strong due 
to the reduced dimensionality of the system. Solving the effective-mass 
Bethe–Salpeter equation for spinless excitons, Ando \cite{Ando1997} 
(and later Hartmann and coworkers \cite{Hartmann2011}) found that 
electron-hole Coulomb attraction is below the critical threshold for 
the transition to the excitonic insulator. Later, Rontani \cite{Rontani2014} 
suggested that intervalley exchange interaction affecting triplet excitons 
may lead to an excitonic instability. This instability does not depend 
on the size of the energy gap, wich is modulated by the axial 
magnetic field, as a consequence of the relativistic nature of 
Dirac fermions. Moreover, the excitonic ground state increases the 
quasiparticle magnetization, which could shed some light on recent 
experiments in ultraclean devices \cite{Laird2015}.

\section{Two-band versus BCS model}\label{s:twoband}

In this section we contrast the mean-field theory of the EI to  
the BCS theory of superconductors. 
We compare the equations of two model junctions, 
(i) one between EI and semimetal (SM-EI), (ii)
the other one between superconductor and normal metal (N-S). 
In both cases the phase boundary is
due to the variation of the order parameter 
that changes along the direction perpendicular to the
interface, tending respectively to zero in the bulk normal
phase and to a constant value inside the condensed phase. 
The quasiparticle amplitudes for both SM-EI and N-S junctions are 
formally identical and 
are used in Section \ref{s:Andreev} to compute the flow of charge 
and heat through the junction.   

\subsection{The SM-EI junction}\label{s:SM-EI}

\begin{figure}
\setlength{\unitlength}{1 cm}
\begin{picture}(8.5,9.5)
\put(1.4,-3.2){\epsfig{file=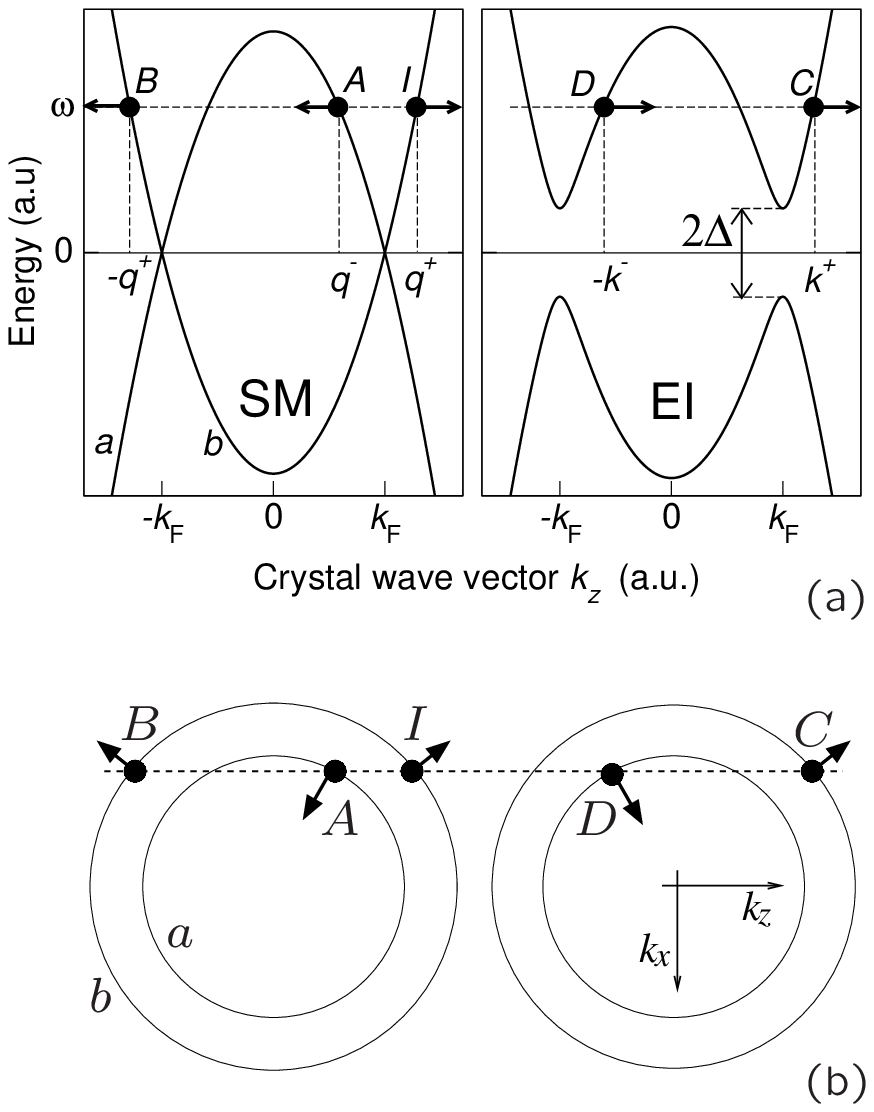,width=4.0in,,angle=0}}
%\put(1.8,1.6){(a)}
%\put(8.0,1.6){(c)}
%\put(5.0,1.6){(b)}
\end{picture}
\caption{Junction between  semimetal (SM, left) and
excitonic insulator (EI, right). (a) Quasiparticle energy $\omega$
vs wave vector $k_z$. The labels mark the
allowed elastic scattering channels for
an incoming electron (labeled $I$) with $k_z = q^+$.
In particular, $A$ is the interband (Andreev)
reflection, $B$ the intraband reflection,
$C$ the intraband transmission, and $D$ the interband transmission.
The size of the EI gap is $2\Delta$.
(b) Isoenergetic contour lines in the 
$(k_x,k_z)$ space for the energy $\omega$
shown in panel a. The arrows point to the
group velocities of electrons in the different scattering channels.
}
\label{figura1}
\end{figure}

\begin{figure}
\setlength{\unitlength}{1 cm}
\begin{picture}(8.5,11.0)
\put(2.1,-1.5){\epsfig{file=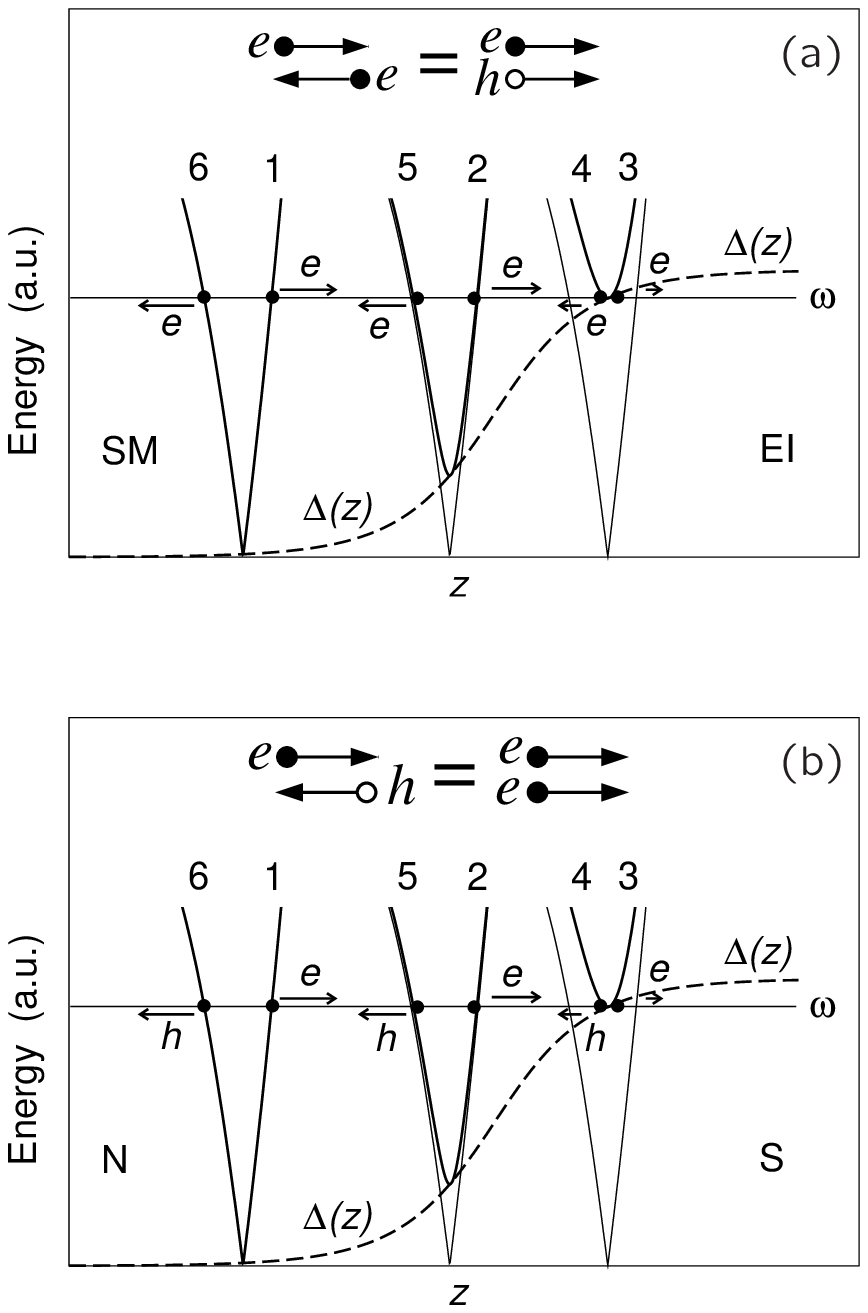,width=3.8in,,angle=0}}
%\put(1.8,1.6){(a)}
%\put(8.0,1.6){(c)}
%\put(5.0,1.6){(b)}
\end{picture}
\caption{Andreev reflection.
(a) SM-EI junction. An incoming $b$-band electron is backscattered
into the $a$ band. The thicker (thinner) curves are
the renormalized (bare) bands at different values of the
$z$ coordinate, with $\Delta(z)$ being the corresponding
order parameter. Only the relevant low-energy portion
of the spectrum is shown here, with numbers from 1 to 6 pointing to 
the time sequence of the reflection process (the arrows represent 
group velocities). The inset illustrates that the reflection process 
may alternatively be seen as a 
coherent flow of excitons from the SM into the EI.
(b) N-S junction. In contrast to panel a, an electron is backscattered
as a hole, hence the whole process may be thought of as the dissipationless
flow of Cooper pairs through the interface.
}
\label{figura2}
\end{figure}

We start studying the junction between semimetal (SM) and EI
on the basis of a spinless two-band model. 
The SM has overlapping isotropic conduction
and valence bands ($b$ and $a$, respectively) of opposite curvature 
and one electron per unit cell, hence 
the Fermi surface is a sphere in momentum space,
located at the nesting of the two
bands; since there are $N$ electrons and $2N$ states available,
the nesting occurs at zero energy in Fig.~\ref{figura1}.
One may turn the SM into an EI by either changing the SM stoichiometric
composition through suitable alloying or applying stress, which 
opens a gap of size $2\Delta$ in the bulk EI in virtue of the strong
inter-band Coulomb interaction [cf.~right panel of Fig.~\ref{figura1}(a)].
The variation of the EI order parameter $\Delta(\bm{r})$ 
(defined below) along the coordinate $z$ normal to the interface 
determines the effective interface potential,
as shown in Fig.~\ref{figura2}(a).

The Hamiltonian of the SM-EI junction is
\begin{equation}
\mathcal{H}_{\text{SM-EI}}=\mathcal{H}_{0}+\mathcal{H}_{1}+\mathcal{H}_{2}.
\label{eq:H_EISM}
\end{equation}
Here $\mathcal{H}_{1}$ is the kinetic term which embodies the effect
of the ideal and frozen crystal lattice on electrons
with the envelope function in the effective mass approximation
\cite{Kohn1955}:
\begin{equation}
\mathcal{H}_{1}=\sum_{i=a,b}\int \!\!\text{d}\,\bm{r}\, \psi_i^{\dagger}
\!\left(\bm{r}\right)\varepsilon_i\!\left(\bm{r}\right)
\psi_i\!\left(\bm{r}\right).
\label{eq:h1}
\end{equation}
The field operator $\psi_a\!\left(\bm{r}\right)$
[$\psi_b\!\left(\bm{r}\right)$] annihilates an electron in the valence
(conduction) energy band at the position $\bm{r}$ in space.
The real-space band operators $\varepsilon_i\!\left(\bm{r}\right)$
appearing in Eq.~(\ref{eq:h1}) take the form:
\begin{subequations}
\begin{eqnarray}
\varepsilon_a\!\left(\bm{r}\right) &=& G/2+(2m)^{-1}\nabla^2;\\
\varepsilon_b\!\left(\bm{r}\right) &=& -G/2-(2m)^{-1}\nabla^2.
\end{eqnarray}
\end{subequations}
Here $m$ is the (positive) effective mass, $G$ is the (positive) band overlap, 
and the energies are 
measured from the Fermi surface \cite{Kohn1967}.
Throughout this work we put $\hbar=1$ and assume that the system has
unit volume.  
The valence- and conduction-band energy levels of 
the non-interacting bulk crystal, eigenvalues of $\mathcal{H}_{1}$, are 
\begin{subequations}
\label{eq:bandstructure}
\begin{eqnarray}
\varepsilon_a\!\left(\bm{k}_a\right)&=&G/2-(2m)^{-1}k_a^2,\\
\varepsilon_b\!\left(\bm{k}_b\right)&=&-G/2+(2m)^{-1}k_b^2,
\end{eqnarray}
\end{subequations}
where $\bm{k}_a$ and $\bm{k}_b$ refer to the respective band extrema.
We assume the valence-band has
a single maximum at $\bm{k}=0$ whereas the conduction-band a single minimum at
$\bm{k}=\bm{w}$, and ignore complications due to the
presence of equivalent extrema.  
The Fermi wave vector is given by $k_{\text{F}}^2=m G$.
The two-body term $\mathcal{H}_{2}$ consists of
the inter-band Coulomb interaction,
\begin{equation}
\mathcal{H}_{2} =
\int \!\!\text{d}\,\bm{r}\,\text{d}\,\bm{r'}\,
\psi_a^{\dagger}\!\left(\bm{r}\right)
\psi_b^{\dagger}\!\left(\bm{r'}\right)
V_2\!\left(\bm{r-r'}\right)
\psi_b\!\left(\bm{r'}\right)\psi_a\!\left(\bm{r}\right),
\end{equation}
with $V_2\!\left(\bm{r}\right)$ being the dielectrically screened Coulomb
potential \cite{Keldysh1964}. 
Renormalization effects due to
intra-band Coulomb interaction and temperature dependence are 
taken into account into the energy band structure (\ref{eq:bandstructure}).
The one-body term $\mathcal{H}_{0}$ is the sum of two parts,
\begin{equation}
\mathcal{H}_{0}=\mathcal{V} + \mathcal{V}_{\text{hyb}}.
\end{equation}
$\mathcal{V}$ is the intra-band term,
\begin{equation}
\mathcal{V}=\sum_{i=a,b}\int \!\!\text{d}\,\bm{r}\,
\psi_i^{\dagger}\!\left(\bm{r}\right)V\!\left(\bm{r}\right)
\psi_i\!\left(\bm{r}\right),
\end{equation}
which includes the effects of the band offset as well as those of
possible impurities and defects at the
interface, such as a thin insulating layer,
via the single-particle potential
$V\!\left(\bm{r}\right)$. 
The potential $V\!\left(\bm{r}\right)$ can also include
the effect of a voltage bias applied to the junction in a steady-state
regime. The inter-band term,
\begin{equation}
\mathcal{V}_{\text{hyb}}=\int \!\!\text{d}\,\bm{r}\,
\psi_b^{\dagger}\!\left(\bm{r}\right)V_{\text{hyb}}\!\left(\bm{r}
\right) \psi_a\!\left(\bm{r}\right)\quad + \text{\;\,H.c.},
\end{equation}
hybridizes $b$ and $a$ bands by means of
the potential 
$V_{\text{hyb}}\!\left(\bm{r}\right)$. 
This term may be originated e.g.~by 
the change of an element in the SM compound. 
The influence of the potential $V_{\text{hyb}}\!\left(\bm{r}\right)$ 
on exciton condensation, which depends on the
symmetry of the bands involved,
is discussed in Refs.~\cite{Portengen1996a,Portengen1996b,Duan1997}.

Both the hybridization potential and the EI order parameter   
contribute additively to the small band 
gap formed between the two overlapping bands [cf.~right panel 
in Fig.~\ref{figura1}(a)] hence their separate contributions cannot be 
distinguished by spectroscopy, 
as it was early recognized \cite{Kohn1967,Kohn1970,Guseinov1973}.
However, in this Chapter we show that the transport properties across  
the SM-EI junction are distinctive features of the EI.  
The key point is in the length scale of the variation of 
the effective interface potential which reflects or transmits 
the electron. A component of the effective potential is the 
position dependent order parameter
$\Delta(\mathbf{r})$, which decreases from the 
bulk value in the EI region to zero in the SM region [see 
Fig.~\ref{figura2}(a)]. The length scale of the change is the coherence 
length in the EI, much longer than the lattice constant. 
Since this scenario is most likely if the lattices of the 
two components are as similar as possible, we classify as homogeneous
the junction with $\Delta(\mathbf{r})$ being the only contribution
to the interface potential. 

On the other hand, the one-electron interface 
potential $V_{\text{hyb}}\!\left(\bm{r}\right)$ 
due to the change in hybridization has 
an abrupt variation on a length scale of the order of a few atomic layers,
thus considered in our context as a heterogeneous junction
(this includes the case of a Schottky barrier).
We have studied the common physical features of the heterogeneous junction, 
including the abrupt band edge discontinuity, the short-ranged 
interface potential, and the impurities at the interface. 
Whereas the charge carriers in the heterogeneous junction experience 
uninteresting intraband reflection, 
electrons in the homogeneous junction change valley when backscattered  
as a feature of the EI band mixing. 
In the rest of the Chapter we focus on the homogeneous junction
and refer the reader elsewhere \cite{Rontani2005a} 
for the study of the heterogeneous junction
(corresponding to the case $V_{\text{hyb}}\!\left(\bm{r}\right)\neq 0$).

Following Sham and Rice \cite{Sham1966}, we introduce the electron
quasi-particle amplitudes,
\begin{subequations}
\begin{eqnarray}
f\!\left(\bm{r},t\right) &=& \langle\Psi_0|\tilde{\psi}_b\!\left(\bm{r},t
\right)\!|\Psi_{\bm{k}}^e\rangle,\\
g\!\left(\bm{r},t\right) &=& \langle\Psi_0|\tilde{\psi}_a\!\left(\bm{r},t
\right)\!|\Psi_{\bm{k}}^e\rangle.
\end{eqnarray}
\label{eq:fgdef}
\end{subequations}
Here $|\Psi_0\rangle$ and $|\Psi_{\bm{k}}^e\rangle$ are 
the exact interacting ground states with $N$ and $N+1$
electrons, respectively; the 
quantum index $\bm{k}$ labeling the electron quasiparticle means 
the crystal momentum only in the bulk phase
as the overall translational symmetry is destroyed
by the presence of the junction. 
States and operators are written
in the Heisenberg representation \cite{Abrikosov1975}
(flagged by the tilde symbol on operators):
\begin{equation}
\tilde{\psi}_i\!\left(\bm{r},t\right) =
\exp\!\left(\text{i}\left[\mathcal{H}
-\mu\, \mathcal{N}\right]t\right) 
\psi_i\!\left(\bm{r}\right)
\exp\!\left(-\text{i}\left[\mathcal{H}
-\mu\, \mathcal{N}\right]t\right),
\label{eq:defop}
\end{equation}
where $\mu$ is the chemical potential (here
$\mu=0$ due to electron-hole symmetry) and
the number operator $\mathcal{N}$ is defined by
\begin{equation}
\mathcal{N}=\sum_{i=a,b}\int \!\!\text{d}\,\bm{r}\,  
\psi_i^{\dagger} \!\left(\bm{r}\right) \psi_i\!\left(\bm{r}\right).
\label{eq:numop}
\end{equation}
%\begin{widetext}
Writing down the Heisenberg equations of motion for the operators
$\tilde{\psi}_i\!\left(\bm{r},t\right)$, exploiting 
the mean-field approximation to express them in a closed form, 
and neglecting unessential intra-band Hartree terms,
we derive a set of two coupled
integro-differential equations for the amplitudes
$f\!\left(\bm{r},t\right)$ and $g\!\left(\bm{r},t\right)$:
\begin{subequations}
\label{eq:BdG}
\begin{eqnarray}
&& \text{i}\frac{\partial f\!\left(\bm{r},t\right)}{\partial t} =
\left[ -\frac{\nabla^2}{2m} -\frac{k_{\text{F}}^2}{2m} 
+ V\!\left(\bm{r}\right)
\right]f\!\left(\bm{r},t\right) 
+\!\!\int \!\!\text{d}\bm{r'}
\Delta\!\left(\bm{r},\bm{r'}\right)
g\!\left(\bm{r'},t\right),\label{eq:BdGa} \\
&& \text{i}\frac{\partial g\!\left(\bm{r},t\right)}{\partial t} =
\left[\frac{\nabla^2}{2m} +\frac{k_{\text{F}}^2}{2m} 
+ V\!\left(\bm{r}\right)
\right]g\!\left(\bm{r},t\right) 
+\!\!\int \!\!\text{d}\bm{r'}
\Delta^*\!\left(\bm{r'},\bm{r}\right)
f\!\left(\bm{r'},t\right).\label{eq:BdGb}
\end{eqnarray}
\end{subequations}

The built-in coherence of the exciton condensate,
$\Delta\!\left(\bm{r},\bm{r'}\right)$, appearing in Eqs.~(\ref{eq:BdG})
for $k> k_{\text{F}}$ is defined as
\begin{equation}
\Delta\!\left(\bm{r},\bm{r'}\right)=V_2\!\left(\bm{r-r'}\right)
\langle\Psi_0|\tilde{\psi}_b\!\left(\bm{r}\right)
\tilde{\psi}_a^{\dagger}\!\left(\bm{r'}\right)|\Psi_0\rangle.
\label{eq:delta}
\end{equation}
Except for the factor $V_2\!\left(\bm{r-r'}\right)$,
$\Delta\!\left(\bm{r},\bm{r'}\right)$
is the exciton macroscopic wave function.
In fact, $\langle\Psi_0|\tilde{\psi}_b\!\left(\bm{r}\right)
\tilde{\psi}_a^{\dagger}\!\left(\bm{r'}\right)|\Psi_0\rangle$
is the average on the many-electron ground-state 
of the operator destroying an electron-hole pair, i.e.,
one $b$-band electron at $\bm{r}$ and one $a$-band hole at $\bm{r'}$
[the electron creation operator $\tilde{\psi}_a^{\dagger}\!
\left(\bm{r'}\right)$ may be regarded as a hole destruction operator].
Such average is zero in the SM phase, since 
for $k> k_{\text{F}}$ $b$-band
levels are empty and $a$-band levels filled, but it acquires a finite
value in the EI phase. 
Besides, the finiteness of $\Delta\!\left(\bm{r},\bm{r'}\right)$
reflects the new periodicity in real space of the EI phase,
as the electron density shows an additional charge-density-wave-like order 
characterized by the wave vector
$\bm{w}$ displacing the extrema of $a$ and $b$ bands \cite{Kohn1968}.
For $k< k_{\text{F}}$ the roles of electrons and holes are exchanged
hence the definition (\ref{eq:delta})
of $\Delta\!\left(\bm{r},\bm{r'}\right)$
is modified accordingly.

The built-in coherence $\Delta\!\left(\bm{r},\bm{r'}\right)$ 
generically depends on both center-of-mass
and relative-motion coordinates, but inside the EI bulk  
the center-of-mass part of the condensate wave function is a plane 
wave with zero momentum, hence 
$\Delta$ depends only on the relative coordinate $\bm{r-r'}$.
We expect $\Delta(\bm{r-r'})$ to
smoothly vanish when $\left|\bm{r-r'}\right|$ becomes larger than 
the characteristic length, the exciton radius.
This allows us to easily find the bulk solution of the system of eqns
(\ref{eq:BdGa}-\ref{eq:BdGb}) 
[$V\!\left(\bm{r}\right) =0$] 
in terms of the two-component plane wave
\begin{equation}
{f_{\bm{k}}\!\left(\bm{r},t\right) \choose
g_{\bm{k}}\!\left(\bm{r},t\right)} =
{ u_{\bm{k}} \choose v_{\bm{k}} } {\rm e}^{ \text{i}\left(\bm{k\cdot r}
-\omega t \right) },
\label{eq:bulk}
\end{equation}
with energy
\begin{equation}
\omega\!\left(\bm{k}\right)=
\sqrt{\xi_{\bm{k}}^2+\left|\Delta_{\bm{k}}\right|^2},
\label{eq:dispersiong}
\end{equation}
where 
\protect{$\xi_{\bm{k}}=\left( k^2 -k_{\text{F}}^2 \right)/(2m)$} and
$\Delta_{\bm{k}}$ is the Fourier component of $\Delta\!\left(\bm{r}\right)$.
The amplitudes are normalized as
\begin{equation}
\left|u_{\bm{k}}\right|^2=\frac{1}{2}\!
\left(1+\frac{\xi_{\bm{k}}}{E_{\bm{k}}}
\right),\qquad \left|u_{\bm{k}}\right|^2+\left|v_{\bm{k}}\right|^2=1,
\end{equation}
and the relative phase between $u_{\bm{k}}$ and $v_{\bm{k}}$ is given by
\begin{equation}
\frac{u_{\bm{k}}}{v_{\bm{k}}}=
\frac{\Delta_{\bm{k}}}{\omega\!\left(\bm{k}\right)-\xi_{\bm{k}}}.
\end{equation}
When $\Delta_{\bm{k}}=0$, the amplitude (\ref{eq:bulk}) is
the trivial solution with $u_{\bm{k}}=1$ and
$v_{\bm{k}}=0$, i.e., a conduction-band plane wave. 
When excitons form a condensate, solution
(\ref{eq:bulk}) is admissible
only if the self-consistency condition derived by the definition
of $\Delta\!\left(\bm{r}\right)$ is satisfied. This condition, which
can be easily obtained from Eq.~(\ref{eq:delta}), is formally analogous
to the BCS gap equation:
\begin{equation}
\Delta_{\bm{k}}=\sum_{\bm{p}}
\frac{V_{2,\,\bm{k-p}}
\Delta_{\bm{p}}}{2\omega\!\left(\bm{p}\right)},
\label{eq:gap}
\end{equation}
with $V_{2,\,\bm{k}}$ being the Fourier component of
$V_2\!\left(\bm{r}\right)$.

In general, 
the amplitudes $f\!\left(\bm{r},t\right)$ and $g\!\left(\bm{r},t\right)$
are the position space representation of the stationary electron-like
elementary excitation across the {\em whole} junction. Taken
together, they signify the wave function of the quasiparticle: $f$ ($g$)
is the probability amplitude for an electron
of belonging to the conduction (valence) band. They satisfy the
normalization condition
\begin{equation}
\int \!\!\text{d}\,\bm{r}\,\left[\left|f\!\left(\bm{r},t\right)\right|^2
+\left|g\!\left(\bm{r},t\right)\right|^2\right]=1,
\end{equation}
and have always positive excitation energy $\omega$ due to the
definitions (\ref{eq:fgdef}-\ref{eq:defop}).
The probability current density
$\bm{J}\!\left(\bm{r},t\right)$ can be found starting from the
definition of the probability density
$\rho\!\left(\bm{r},t\right)$ for finding either a conduction-
or a valence-band electron at a particular time and place,
\protect{$\rho\!\left(\bm{r},t\right)=
\left|f\right|^2 +\left|g\right|^2$}. After some manipulation of
the equations of motion (\ref{eq:BdG}), one derives the continuity
equation
\begin{equation}
\frac{\partial \rho}{\partial t}+\nabla\cdot {\bm{J}}=0,
\label{eq:continuity}
\end{equation}
where
\begin{equation}
\bm{J}={\text{Im}}\!\left\{
f^*\frac{\nabla}{m} f - g^*\frac{\nabla}{m} g\right\}.
\label{eq:J}
\end{equation}
Note that the two terms appearing in the rhs of Eq.~(\ref{eq:J}), referring
respectively to conduction and valence band electrons, have
opposite sign since the curvature of the two bands is opposite.
One can verify that the semiclassical group velocity of the quasiparticle,
\protect{$\bm{v}_{\text{g}}=\nabla_{\bm{k}}\, \omega $},
coincides with the velocity
$\bm{v}$ given by the full quantum mechanical expression (\ref{eq:J}),
with \protect{$\bm{J}=\rho\,\bm{v}$}.

\subsection{The N-S junction}

In this section we highlight the suggestive parallelism between the
formalism introduced in section (\ref{s:SM-EI}) and the treatment 
of quasiparticle excitations in conventional superconductors,
as modeled by the BCS theory.

The Hamiltonian of the N-S junction is
\begin{equation}
\mathcal{H}_{\text{N-S}}=\mathcal{H}^{\prime}_{0}
+ \mathcal{H}^{\prime}_{1}
+ \mathcal{H}^{\prime}_{2}.
\label{eq:H_NS}
\end{equation}
Electrons in the metal experience the crystal lattice potential
through $\mathcal{H}^{\prime}_{1}$, 
\begin{equation}
\mathcal{H}^{\prime}_{1}
=
-\frac{1}{2m}\sum_{\sigma=\uparrow,\downarrow}
\int \!\!\text{d}\,\bm{r}\, 
\psi_{\sigma}^{\dagger}
\!\left(\bm{r}\right)
\nabla^2 
\psi_{\sigma} \!\left(\bm{r}\right) .
\label{eq:hprime1}
\end{equation}
The space-dependent field operator $\psi_{\sigma}  \!\left(\bm{r}\right)$
annihilates an electron with spin $\sigma$ 
in the two-fold degenerate
conduction energy band, whose energy dispersion is taken to be parabolic
for simplicity,
\begin{equation}
\varepsilon\!\left(\bm{k}\right)= k^2/(2m),
\label{eq:bandstructureBCS}
\end{equation}
$\varepsilon\!\left(\bm{k}\right)$ being the eigenvalue 
of $\mathcal{H}^{\prime}_{1}$.
With respect to the SM, the role of $a$ and $b$ bands
is replaced by the two spin flavors $\uparrow$ and $\downarrow$.
The Fermi wave vector $k_{\text{F}}$ is fixed by the condition that
there are $N$ electrons in the system,
with $\mu = k_{\text{F}}^2 / 2m$. 

The relevant two-body interaction $\mathcal{H}^{\prime}_{2}$ 
is attractive and short-ranged,
\begin{equation}
\mathcal{H}^{\prime}_{2} = -g
\int \!\!\text{d}\,\bm{r}\,
\psi_{\uparrow}^{\dagger}\!\left(\bm{r}\right)
\psi_{\downarrow}^{\dagger}\!\left(\bm{r}\right)
\psi_{\downarrow}\!\left(\bm{r}\right)\psi_{\uparrow}\!\left(\bm{r}\right),
\label{eq:effective}
\end{equation}
with $g$ being a positive constant parametrizing the 
combined effect of Coulomb and electron-phonon
interaction in the vicinity of the Fermi surface \cite{deGennes1999}.
The short-range interaction (\ref{eq:effective}) does not affect
electrons with parallel spin
as a consequence of Pauli exclusion principle.
The effective potential (\ref{eq:effective}) 
results from the competition between Coulomb repulsion and
the screening effect of the positive ions in the lattice. Close to the 
resonance frequency of the ion motion, the ions give a very large response
to the perturbation induced by an electron charge. The resulting 
cloud of the moving electron plus the polarized ions has a net positive charge,
then inducing a weak electron-electron attraction, whose characteristic
energy is a tiny fraction of the Fermi energy.
As in the case of the SM-EI junction, the boundary between N and S phases
is determined by the variation along $z$
of the pair potential associated to $\mathcal{H}^{\prime}_{2}$,
defined below [see Fig.~\ref{figura2}(b)].
A residual effect of 
Coulomb interaction is to shift the energy levels, that 
are already renormalized in the dispersion relation
(\ref{eq:bandstructureBCS}). 
The one-body term 
\begin{equation}
\mathcal{H}^{\prime}_{0}=\sum_{\sigma=\uparrow,\downarrow}
\int \!\!\text{d}\,\bm{r}\,
\psi_{\sigma}^{\dagger}\!\left(\bm{r}\right)V^{\prime}\!\left(\bm{r}\right)
\psi_{\sigma}\!\left(\bm{r}\right),
\end{equation}
arises from the 
possible impurities and defects at the
interface, as well as the applied bias voltage.

In order to find out the quasiparticles of the N-S junction,
we follow the same approach
as for the SM-EI junction, with one important difference \cite{Andreev1964}
that derives from the following
definition of the amplitudes:
\begin{subequations}
\begin{eqnarray}
f^{\prime}\!\left(\bm{r},t\right) &=& \langle\Psi_0^{\prime}|
\tilde{\psi}_{\uparrow}\!\left(\bm{r},t
\right)\!|\Psi_{\bm{k}}^b\rangle,\label{eq:fgdef_bis_a}\\
g^{\prime}\!\left(\bm{r},t\right) &=& \langle\Psi_0^{\prime}|
\tilde{\psi}^{\dagger}_{\downarrow}\!\left(\bm{r},t
\right)\!|\Psi_{\bm{k}}^b\rangle. \label{eq:fgdef_bis_b}
\end{eqnarray}
\label{eq:fgdef_bis}
\end{subequations}
Here $|\Psi_{\bm{k}}^b\rangle$ is
the state with one quasiparticle
added to the many-electron ground state $|\Psi_0^{\prime}\rangle$.
According to eqns (\ref{eq:fgdef_bis}), the number of
particles is not a constant of motion, as we 
add both an electron [eqn (\ref{eq:fgdef_bis_a})]
and a hole [eqn (\ref{eq:fgdef_bis_b})] to $|\Psi_0^{\prime}\rangle$.
This is allowed within the gran canonical framework, where
the chemical potential $\mu$ is the independent thermodynamic 
variable instead of $N$. 

The resulting `Bogoliubov-de Gennes' equations
of motion are:
\begin{subequations}
\label{eq:BdG2}
\begin{eqnarray}
&& \text{i}\frac{\partial f^{\prime}\!\left(\bm{r},t\right)}{\partial t} =
\left[ -\frac{\nabla^2}{2m} -\frac{k_{\text{F}}^2}{2m} 
+ V^{\prime}\!\left(\bm{r}\right)
\right]f^{\prime}\!\left(\bm{r},t\right) 
+ \Delta^{\prime}\!\left(\bm{r}\right)
g^{\prime}\!\left(\bm{r},t\right),\label{eq:BdGa2} \\
&& \text{i}\frac{\partial g^{\prime}\!\left(\bm{r},t\right)}{\partial t} =
\left[\frac{\nabla^2}{2m} +\frac{k_{\text{F}}^2}{2m} 
- V^{\prime}\!\left(\bm{r}\right)
\right]g^{\prime}\!\left(\bm{r},t\right) 
+ \Delta^{\prime *}\!\left(\bm{r}\right)
f^{\prime}\!\left(\bm{r},t\right),\label{eq:BdGb2}
\end{eqnarray}
\end{subequations}
with the local pair potential $\Delta^{\prime}\!\left(\bm{r}\right)$
being defined as
\begin{equation}
\Delta^{\prime}\!\left(\bm{r}\right)=-g
\langle\Psi_0^{\prime}|\tilde{\psi}_{\downarrow}\!\left(\bm{r}\right)
\tilde{\psi}_{\uparrow}\!\left(\bm{r}\right)|\Psi_0^{\prime}\rangle.
\label{eq:delta2}
\end{equation}
The space-dependent parameter $\Delta^{\prime}\!\left(\bm{r}\right)$
may be regarded as the center-of-mass wave function of the condensate
made of Cooper pairs. The latter are bound pairs of two electrons 
with opposite spins, as it is evident from the definition (\ref{eq:delta2}).
The product of the two operators that destroys a Cooper pair, appearing in the
so-called anomalous average (\ref{eq:delta2}),
does not commute with the number operator $\mathcal{N}$ as it
breaks the $U$(1) gauge symmetry
of total Hamiltonian $\mathcal{H}_{\text{N-S}}$. 
Nevertheless, since the number of electrons is macroscopic, 
the number fluctuations, of order $\sqrt{N}$, are small and may be neglected
with respect to the average value of $N$. Moreover,
since a value of one or two is still small with respect to $\sqrt{N}$,
the ground states with either $N$ or $N-2$ electrons 
must be regarded as identical,
so the anomalous character (i.e., breaking the gauge symmetry
of $\mathcal{H}_{\text{N-S}}$) of the definitions
(\ref{eq:delta2}) and
(\ref{eq:fgdef_bis}) is physically irrelevant.

It is remarkable to observe that the systems of equations (\ref{eq:BdG}) and 
(\ref{eq:BdG2}), respectively describing the SM-EI and the N-S
junctions, are formally identical in the homogeneous case.
This corresponds to put
respectively $ V\!\left(\bm{r}\right)=0$ in eqns
(\ref{eq:BdG}) and $ V^{\prime}\!\left(\bm{r}\right)=0$ in eqns
(\ref{eq:BdG2}), as well as to take 
the built-in exciton coherence in (\ref{eq:BdG}) as a local operator, 
$\Delta\!\left(\bm{r},\bm{r'}\right)=\delta\!\left(\bm{r-r'}\right)
\Delta\!\left(\bm{r}\right)$ (then $\Delta_{\bm{k}}$ does not depend 
on $\bm{k}$).
However, the quasiparticle amplitudes for the two model junctions
signify profoundly different types of single-particle excitations.

In the EI, to obtain a free electron
in the $b$-band---for $k>k_{\text{F}}$---one 
has to break an exciton among those forming the 
condensate, that is
a  bound pair of a $b$-band electron and $a$-band hole. 
The way to do this is to either
create an electron in the $b$ band, whose amplitude component is $f$, or
destroy a hole in the $a$ band, whose amplitude component is $g$.  

In the S, to obtain an unbounded single particle with spin $\uparrow$,
one has to break a Cooper pair in the condensate. This is
accomplished by means of either
creating an electron with spin $\uparrow$ or destroying an electron with spin
$\downarrow$. The latter option is equivalent to creating a 
hole with spin $\uparrow$,
as a consequence of time-reversal symmetry.
The components $f^{\prime}$ and $g^{\prime}$ are the amplitudes for
the propagation of the electron and the hole, respectively.
Table \ref{t:features} compares the key features of the EI with those of 
the S, with regards to both the ground state and the quasiparticle excitations. 

In the following we are interested in comparing the SM-EI
and N-S systems. In order to stress their formal analogy, hereafter we 
drop the prime symbol to label quantities referring to the N-S junction
(as $f$, $g$, $\Delta$, etc) and use the same notation in both cases. 
With this convention, formulae (\ref{eq:bulk}--\ref{eq:J}) obtained
for the SM-EI junction hold for the N-S junction, too.

\section{Andreev reflection at the interface
between excitonic insulator and semimetal}\label{s:Andreev} 

In this section we introduce the phenomenon of Andreev reflection
at the N-S boundary as a paradigm to discuss the transport through
the SM-EI junction. 
There are three qualitatively important results that are common to
both systems:
(i) all three cartesian
components of the velocity change sign when the quasiparticle is reflected
(Sec.~\ref{s:velocity})
(ii) the ratio of incident quasiparticles
$C(\omega)$ which are transmitted
through the interface depends on the coherence factors
of the condensate, being strongly suppressed close to the gap
(Sec.~\ref{s:transmission})
(iii) the condensate induces pairing on the normal side of the
junction (proximity effect, Sec.~\ref{s:proximity}).

\subsection{Velocity inversion at the interface}\label{s:velocity}

\begin{figure}
\setlength{\unitlength}{1 cm}
\begin{picture}(8.5,7.0)
\put(2.0,-0.0){\epsfig{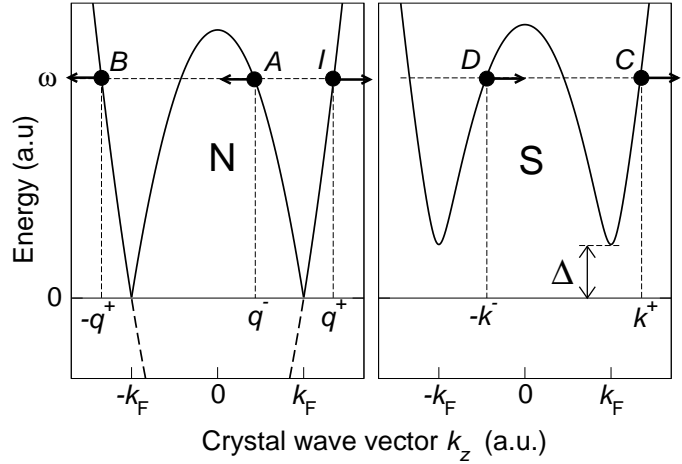}}
%\put(1.8,1.6){(a)}
%\put(8.0,1.6){(c)}
%\put(5.0,1.6){(b)}
\end{picture}
\caption{Junction between  normal metal (N, left) and
superconductor (S, right). The plot shows the quasiparticle energy $\omega$
vs wave vector $k_z$ in the two bulk phases. The labels mark the
allowed elastic scattering channels for
an incoming particle (labeled $I$) with $k_z = q^+$.
$A$ is the Andreev reflection, $B$ the normal reflection,
$C$ the normal transmission, and $D$ the cross-branch transmission.
Note that the energy of the particle is positive: on the N side the 
hole branch for $ \left|k_z\right| < k_{\text{F}}$ is obtained by 
inverting the energy of the portion of band filled with electrons
(showed as a dashed curve) with respect to the Fermi surface.
}
\label{BCSdisp}
\end{figure}

The electrical transport across the
N-S junction exhibits high conductance behavior at vanishing applied
voltage bias. This evidence seems to contradict 
the fact that quasiparticle excitations are gapped in the S
(see Fig.~\ref{BCSdisp}): quasiparticles 
in the bulk N approaching the junction  
with energy smaller than the gap, $0< \omega < \Delta$,
cannot penetrate into the bulk S. This paradox is solved  
by Andreev reflection, which is illustrated below.

Consider $\Delta\!\left(z\right)$ to be a smooth complex
increasing function of $z$, tending respectively to the asymptotic values
zero when $z\rightarrow -\infty$, inside the bulk N, and 
$\Delta_0$ when $z\rightarrow +\infty$, inside the bulk S
[Fig.~\ref{figura2}(b)]. 
Following Andreev \cite{Andreev1964},
we note that the medium under consideration is completely homogeneous 
with an accuracy $2m\left|\Delta_0\right|/k_{\text{F}}^2$---a very
small quantity in typical superconductors.
Therefore, 
we seek a solution of eqns (\ref{eq:BdG2}) in the form
\begin{equation}
f\!\left(\bm{r}\right)={\text{e}}^{\text{i}k_{\text{F}}\bm{n\cdot r}}
\eta\!\left(\bm{r}\right),\qquad
g\!\left(\bm{r}\right)={\text{e}}^{\text{i}k_{\text{F}}\bm{n\cdot r}}
\chi\!\left(\bm{r}\right),
\label{eq:ansatz}
\end{equation}
where $\bm{n}$ is some unit vector and $\eta\!\left(\bm{r}\right)$
and $\chi\!\left(\bm{r}\right)$ are functions that vary slowly
compared to ${\text{e}}^{\text{i}k_{\text{F}}\bm{n\cdot r}}$.
Substituting (\ref{eq:ansatz}) in (\ref{eq:BdG2}) and
neglecting higher derivatives of $\eta$ and $\chi$, we obtain
\begin{subequations}
\label{eq:BdGAndreev}
\begin{eqnarray}
\left(\text{i}v_{\text{F}}\bm{n}\cdot\nabla+\omega\right)
\eta\!\left(\bm{r}\right)-\Delta\!\left(z\right)\chi\!\left(\bm{r}\right)
&=& 0,\\
\left(\text{i}v_{\text{F}}\bm{n}\cdot\nabla-\omega\right)
\chi\!\left(\bm{r}\right)+\Delta^*\!\left(z\right)
\eta\!\left(\bm{r}\right) &=& 0,
\end{eqnarray}
\end{subequations}
where $v_{\text{F}}=k_{\text{F}}/m$.
It is easy to find for $z\rightarrow\pm\infty$ the asymptotic form of the
solutions of eqns (\ref{eq:BdGAndreev}) describing the reflection of
the quasiparticle falling on the junction.
When $z\rightarrow-\infty$ we put $\Delta\!\left(z\right)=0$.
Then
\begin{equation}
{ \eta \choose \chi } = C_1 { 1 \choose 0 } {\text{e}}^{\text{i}
\bm{k}_1\bm{\cdot r}} + C_2 { 0 \choose 1 } {\text{e}}^{\text{i}
\bm{k}_2\bm{\cdot r}},
\label{eq:leftsolution}
\end{equation}
where ${\bm{n\cdot k}_1}=\omega/v_{\text{F}}$,
${\bm{n\cdot k}_2}=-\omega/v_{\text{F}}$; $C_1$ and $C_2$ are arbitrary
constants. The first term on the rhs of eqn (\ref{eq:leftsolution})
corresponds to an electron whose velocity $\bm{v}$ (or $\bm{J}$) 
lies along $\bm{n}$, and the second
term to a hole whose velocity lies in the opposite
direction to $\bm{n}$ (in fact $\omega/v_{\text{F}}\ll k_{\text{F}}$
since \protect{$2m\left|\Delta_0\right|/k_{\text{F}}^2 \ll 1$}).
If $n_z>0$, then the wave function (\ref{eq:leftsolution}) describes an
electron incident on the boundary and
reflected as a hole on the N side;
if $n_z<0$, it describes an incident
hole reflected as an electron.
When $z\rightarrow +\infty$ we put 
\protect{$\Delta\!\left(z\right)=\Delta_0$} in
eqn (\ref{eq:BdGAndreev}). The solution
describing the transmitted wave ($J_z>0$) has for
$\omega>\left|\Delta_0\right|$ the form
\begin{equation}
{ \eta \choose \chi } = \frac{C_3}{\sqrt{2}}
{
\sqrt{1 + v_{\text{F}}\bm{n\cdot k}_3/\omega}
\,{\text{e}}^{\text{i}\varphi/2}
\choose
\sqrt{1 - v_{\text{F}}\bm{n\cdot k}_3/\omega}
\,{\text{e}}^{-\text{i}\varphi/2}
}
{\text{e}}^{ \text{i}\bm{k}_3\bm{\cdot r} },
\label{eq:rightsolution}
\end{equation}
where $C_3$ is a constant, $\varphi$ is the phase of 
$\Delta_0$,
\begin{subequations}
\begin{eqnarray}
{\bm{n\cdot k}_3}&=& v_{\text{F}}^{-1}
\sqrt{\omega^2-\left|\Delta_0\right|^2}\quad\text{for}\quad n_z
>  0,\\
{\bm{n\cdot k}_3}&=&- v_{\text{F}}^{-1}
\sqrt{\omega^2-\left|\Delta_0\right|^2}\quad\text{for}\quad n_z
< 0.
\end{eqnarray}
\end{subequations}

\begin{figure}
\setlength{\unitlength}{1 cm}
\begin{picture}(8.5,4.0)
\put(0.5,-7.0){\epsfig{file=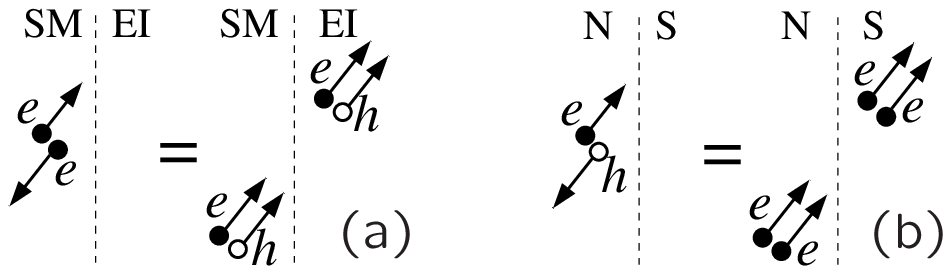,width=5.0in,,angle=0}}
%\put(1.8,1.6){(a)}
%\put(8.0,1.6){(c)}
%\put(5.0,1.6){(b)}
\end{picture}
\caption{Sketch of Andreev reflection for quasiparticles approaching
the junction from the normal-phase side. (a) SM-EI junction. 
The reflected left-going electron is equivalent to a right-going hole. 
(b) N-S junction.
The reflected left-going hole is equivalent to a right-going electron.
}
\label{equivalence}
\end{figure}

As expected, for $\omega<\left|\Delta_0\right|$ the functions $\eta$ and
$\chi$ decay exponentially as $z\rightarrow +\infty$, hence the 
quasiparticle is prevented from entering the bulk S. However,
all three cartesian components of the velocity of the reflected
particle change sign
[see Fig.~\ref{equivalence}(b)].
This remarkable behavior, which is not due to
interface roughness since we take the interface to be completely flat, is the
key to explain the electric transport through the N-S junction. 
An electron with velocity $\bm{v}$ is 
Andreev-reflected into a hole with velocity 
$-\bm{v}$ which carries exactly the same current as the incident electron.
In fact, in virtue of time-reversal invariance, the hole moving with velocity 
$-\bm{v}$ may be regarded as an electron moving with velocity $\bm{v}$
[Fig.~\ref{equivalence}(b)]. 
Therefore, we may understand the process of Andreev reflection as 
an electron above the Fermi surface forming a Cooper pair with another electron below the Fermi surface
on the N side: such pair moves to the S side merging into the condensate, 
whereas the second electron leaves a hole behind 
in the N Fermi sea \cite{Zagoskin1998}. 

The results obtained in this section
hold also for the SM-EI junction, provided one links $\eta$ and $\chi$
components to the probability amplitudes of an electron of being in either $b$ 
or $a$ band, respectively,
as illustrated in Fig.~\ref{equivalence}(a). 
Apparently, the reflection process at the SM-EI junction
seems the usual reflection of an electron from the gap barrier.
However, the complete reversal of the velocity vector
suggests that the reflected electron may be regarded 
as an incoming hole with the same velocity as the incoming electron.
The idea is that the overall reflection process may be thought of as the flow 
of electron-hole pairs---excitons---from the SM to the EI side,
where they merge into the exciton condensate.
Below we substantiate this alternate interpretation. 

\subsection{Coherence factors in the transmission 
coefficients}\label{s:transmission}

To proceed we specify the functional form of the interface potential,
assuming that the excitonic coherence is 
a step function at the SM-EI interface,
\protect{$\Delta\!\left(\bm{r}\right) = \Delta\, \theta\!\left(z\right)$},
with $\Delta > 0$. Moreover, we introduce a simple-minded model for the
effect of disorder at the origin (e.g. an insulating layer) through 
the $\delta$-potential $V\!\left(\bm{r}\right) = H\,\delta(z)$.
In the following we abandon the slowly varying amplitude approximation
and look for solutions of the full eqns (\ref{eq:BdG2}), requiring a 
larger number of scattering channels
than those used in Sec.~\ref{s:velocity}.

Carriers 
coming from the bulk SM with energies slightly outside the EI gap 
have, say for the incident electron at $I$, two reflection channels, 
$A$ and $B$, and two transmission channels at $C$ and $D$ (see 
Fig.~\ref{figura1}). 
If the energy lies within the gap, only the two 
reflection channels are possible. Whereas the interface---by breaking the 
lattice translational symmetry---can in principle connect different 
parts of the Brillouin zone \cite{Sham1979},  
here the relevant regions of the wave vector space are the two valleys near 
the gaps in the bulk EI  
for those states with the same component of the wave vector parallel 
to the interface. 

We consider the elastic scattering at equilibrium, matching wave functions
of the incident ($I$), transmitted ($C$ and $D$), and reflected 
($A$ and $B$) states at the boundary, following 
the approach of Ref.~\cite{Blonder1982}.  
In the bulk EI, there are a pair of magnitudes of
$k$ associated with $\omega$, namely
\begin{equation}
k^{\pm}=\sqrt{2m}\sqrt{ k^2_{\text F}/2m \pm
\left(\omega^2 - {\Delta}^2\right)^{1/2} }.
\label{eq:kvect}
\end{equation}
The total degeneracy of relevant states for each $\omega$ is fourfold:
$\pm k^{\pm}$.
The two states $\pm k^+$
have a dominant conduction-band character, whereas the two states $\pm k^-$
are mainly valence-band states.
Using the notation
\begin{equation}
\Psi(z)=
{f\!\left(z\right) \choose
g\!\left(z\right)}
\end{equation}
the wave functions degenerate in $\omega$ are
\begin{equation}
\Psi_{\pm k^+}=
{ u_0 \choose v_0 } {\rm e}^{ \pm\text{i} k^+z },
\qquad
\Psi_{\pm k^-}=
{ v_0 \choose u_0 } {\rm e}^{ \pm\text{i} k^-z },
\label{eq:bulk1D}
\end{equation}
with the amplitudes $u_0,v_0$ defined as
\begin{equation}
u_0=\sqrt{\frac{1}{2}\left[1+\frac{(\omega^2-{\Delta}^2)^{1/2}}{\omega}
\right]},\quad
v_0=\sqrt{\frac{1}{2}\left[1-\frac{(\omega^2-{\Delta}^2)^{1/2}}{\omega}
\right]},
\end{equation}
possibly extended in the complex manifold.
With regards to the SM bulk,
$\Delta=0$ and the two possible magnitudes of the momentum $q$ reduce to
\protect{$q^{\pm}=[2m( k^2_{\text F}/2m \pm \omega  )]^{1/2}$},
with wave functions
\begin{equation}
\Psi_{\pm q^+}=
{ 1 \choose 0 } {\rm e}^{ \pm\text{i} q^+z },
\qquad
\Psi_{\pm q^-}=
{ 0 \choose 1 } {\rm e}^{ \pm\text{i} q^-z },
\label{eq:SMbulk1D}
\end{equation}
for conduction and valence bands, respectively.
The appropriate boundary conditions 
are: (i) Continuity of
$\Psi$ at $z=0$, so \protect{$\Psi_{\text{EI}}(0)=\Psi_{\text{SM}}(0)
\equiv \Psi(0)$}. (ii) $\left[f_{\text{EI}}'(0)-
f_{\text{SM}}'(0)\right]/(2m)=Hf(0)$ and
$\left[g_{\text{EI}}'(0)-
g_{\text{SM}}'(0)\right]/(2m)=-Hg(0)$, the derivative boundary
conditions appropriate for $\delta$-functions. 
(iii) Incoming (incident),
reflected and transmitted wave directions are defined by their group
velocities, i.e.,
an electron incident from the left is
transmitted with 
$v>0$ and reflected with $v<0$.

If an electron incident on the interface from the SM with energy
$\omega>\Delta$ has wave vector $q^+$, the four outgoing channels, 
with probabilities $A$, $B$, $C$, $D$,
have respectively wave vectors $q^-$, $-q^+$, $k^+$, $-k^-$,
as shown in Fig.~\ref{figura1}.
$C$ is the probability of transmission
through the
interface with a wave vector on the same
(i.e., forward) side of the Fermi surface as $q^+$
(i.e., $q^+\rightarrow k^+$, not
$-k^-$), whereas $D$ gives the probability
of transmission on the back side of the Fermi surface (i.e., $q^+\rightarrow
-k^-$). $B$ is the probability of intra-branch reflection, whereas
$A$ is the probability of Andreev (cross-branch) reflection.
The steady state solution of system (\ref{eq:BdG}) is
\begin{displaymath}
\Psi_{\text{SM}}(z) = \Psi_{\text{inc}}(z)+\Psi_{\text{refl}}(z),
\qquad \Psi_{\text{EI}}(z)=\Psi_{\text{trans}}(z),
\end{displaymath}
where
\begin{eqnarray}
&&\Psi_{\text{inc}}(z) =
{ 1 \choose 0 } {\rm e}^{ \text{i} q^+z }, \quad
\Psi_{\text{refl}}(z)  =
a { 0 \choose 1 } {\rm e}^{ \text{i} q^-z } +
b { 1 \choose 0 } {\rm e}^{ -\text{i} q^+z },\nonumber\\
&&\Psi_{\text{trans}}(z)  =
c { u_0 \choose v_0 } {\rm e}^{ \text{i} k^+z } +
d { v_0 \choose u_0 } {\rm e}^{ -\text{i} k^-z }.
\label{eq:boundary}
\end{eqnarray}
Applying the boundary conditions, we obtain a system of four linear
equations in the four unknowns $a$, $b$, $c$, and $d$, which we solve
at a fixed value for $\omega$. We introduce the
dimensionless barrier strength
\protect{$Z=H/v_{\text{F}}$}, 
and approximate 
$k^+=k^-=q^+=q^-\approx k_{\text{F}}$,
on the basis that the ratio $2m\Delta/k^2_{\text{F}}$ is small.
The quantities $A$, $B$, $C$, $D$, are the ratios of the probability
current densities of the specific transmission or reflection channels
to the current of the incident particle,
e.g.~$A=\left|J_A/J_{\text{inc}}\right|$, and so on.
The conservation of probability requires that
\begin{equation}
A + B + C + D = 1.
\label{eq:coeffsum}
\end{equation}
This result is useful in simplifying expressions for energies below
the gap, $\omega<\Delta$, where there can be no transmitted electrons,
so that $C=D=0$. Then, Eq.~(\ref{eq:coeffsum}) reduces simply to
\protect{$A=1-B$}.

We find
\begin{eqnarray}
a&=& \frac{u_0v_0}{\gamma},\nonumber\\
b&=& \frac{\left(u_0^2-v_0^2\right)Z^2 -\text{i}Z}{\gamma},\nonumber\\
c&=& \frac{u_0\left(1+\text{i}Z\right)}{\gamma},\nonumber\\
d&=& -\frac{\text{i}v_0Z}{\gamma},
\end{eqnarray}
\begin{displaymath}
\gamma=Z^2\left(v_0^2-u_0^2\right)+\left(\text{i}Z+1/2\right)2u_0^2.
\end{displaymath}
The probability coefficients are actually the currents, measured in
units of $v_{\text{F}}$. For example,
\protect{$A=\left|J_A\right|/v_{\text{F}}=\left|a\right|^2$}, and \protect{$D=
\left|d\right|^2/\left|v_0^2-u_0^2\right|$}.
The expression for the energy dependences of $A$, $B$, $C$, and $D$
can be conveniently written in terms of the so-called coherence
factors $u_0$ and $v_0$. The results are given in
Table \ref{tab:table2}.
For convenience, in addition to the general
results we also list the limiting forms of the results for
zero barrier ($Z=0$) and for a strong barrier
[\protect{$Z^2(u_0^2-v_0^2)\gg 1$}], as well as for $\Delta = 0$
(the semimetal case).

\begin{table}[tb]
\tableparts
{ 
\caption{Transmission and reflection coefficients for the SM-EI junction. 
$A$ gives the probability of
Andreev reflection (cross-branch), $B$ of ordinary
reflection, $C$ of transmission without branch crossing, and $D$ of
cross-branch transmission. Here 
$\theta = \omega^2 + 4Z^2\omega^2 +
(1+4Z^4)(\Delta^2-\omega^2)-8Z^3\omega\left(\Delta^2-\omega^2\right)^{1/2}$, 
$\gamma=
Z^2\left(v_0^2-u_0^2\right)
+\left(\text{i}Z+1/2\right)2u_0^2$, and $u_0^2=1-v_0^2=1/2\,[1+(\omega^2
-\Delta^2)^{1/2}/\omega]$.
}
\label{tab:table2} 
}
{ 
\begin{tabular}{ccccc}
\hline
& $A$ & $B$ & $C$ & $D$ \\[1pt]
\hline
No condensate & $0$ &
$\frac{Z^2}{1+Z^2}$ & $\frac{1}{1+Z^2}$ & $0$ \\
General form & & & & \\
$\omega < \Delta$ & $\frac{\Delta^2}{\theta}$ & $1-A$ & $0$ & $0$ \\
$\omega > \Delta$ & $\frac{u_0^2v_0^2}{\left|\gamma\right|^2}$ &
$\frac{\left(u_0^2-v_0^2\right)^2Z^4+Z^2}{\left|\gamma\right|^2}$ &
$\frac{u_0^2\left(u_0^2-v_0^2\right)
\left(1+Z^2\right)}{\left|\gamma\right|^2}$ &
$\frac{v_0^2\left(u_0^2-v_0^2\right)Z^2}{\left|\gamma\right|^2}$ \\
No barrier  & & & & \\
$\omega < \Delta$ & $1$ & $0$ & $0$ & $0$ \\
$\omega > \Delta$ & $\frac{v_0^2}{u_0^2}$ & $0$ &
$\frac{u_0^2-v_0^2}{u_0^2}$ & $0$ \\
Strong barrier & & & & \\
$\omega < \Delta$ &
$\frac{\Delta^2}{4Z^4(\Delta^2-\omega^2)}$
& $1-A$ & $0$ & $0$ \\
$\omega > \Delta$ & $\frac{u_0^2v_0^2}{Z^4(u_0^2-v_0^2)^2}$ &
$1-\frac{1}{Z^2(u_0^2-v_0^2)}$ & $\frac{u_0^2}{Z^2(u_0^2-v_0^2)}$ &
$\frac{v_0^2}{Z^2(u_0^2-v_0^2)}$ \\
\hline
\end{tabular}
}
\end{table}

\begin{figure}
\setlength{\unitlength}{1 cm}
\begin{picture}(8.5,8.0)
\put(1.4,0.0){\epsfig{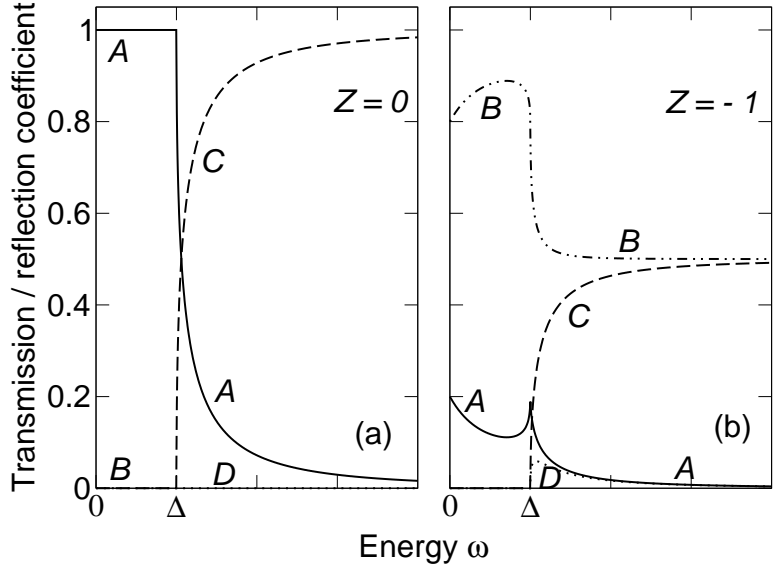}}
%\put(1.8,1.6){(a)}
%\put(8.0,1.6){(c)}
%\put(5.0,1.6){(b)}
\end{picture}
\caption{
Transmission and reflection coefficients at the SM-EI boundary
vs quasiparticle energy $\omega$.
(a) $Z=0$. (b) $Z=-1$.
$A$ gives the probability of Andreev reflection, $B$ gives
the probability of ordinary reflection, $C$ gives the transmission
probability without branch crossing, and $D$ gives the probability
of transmission with branch crossing. The parameter $Z$ measures the
interface transparency.
}
\label{fig5}
\end{figure}

In the absence of disorder ($Z=0$), the dependence
of the transmission coefficient $C(\omega)$ on energies
close to the gap is
\begin{equation}
C(\omega)=2\sqrt{2\left(\omega-\Delta\right)
/\Delta} \qquad
\omega\approx\Delta,
\label{eq:Cdep}
\end{equation}
whereas below the gap the electron is totally Andreev-reflected and
the transmission is zero. The ordinary reflection channel is completely
suppressed ($B=0$) as well as cross-branch transmission ($D=0$).
The situation is depicted in Fig.~\ref{fig5}(a).
Even above the gap, $\omega > \Delta$, there is a high
probability for Andreev reflection, which strongly
depends on $\omega$. For energies close to the gap, $\omega \approx \Delta$,
Andreev reflection is almost certain, $A\approx 1$. 
This is the cause for the low value of interface conductance.
The effect is washed out by the opacity of the interface: as
$\left|Z\right|$ increases [$Z=-1$ in Fig.~\ref{fig5}(b)],
the total reflection probability $A + B$ loses its dependence 
on $\omega$, and the dominant
reflection channel changes from the Andreev one ($A$) into the 
ordinary one ($B$).

The interface opacity $Z$ is the handle to tune the effect
of excitonic coherence on transport, as discussed in the next sections.
Remarkably, the overall set of results of Table \ref{tab:table2}
is formally identical to the analogous quantities 
obtained for the N-S junction
(e.g.~compare with Table II of Ref.~\cite{Blonder1982}), the only slight
difference being the behavior for $Z\neq 0$.
In fact, due to the different boundary conditions, the coefficients
of the N-S junction are even functions of $Z$, whereas
those of Table \ref{tab:table2} do not have a definite parity with 
respect to the sign of $Z$ for $\omega<\Delta$
(there is a mistake in the entry corresponding to the sub-gap 
strong-barrier case 
appearing in Table II of Ref.~\cite{Blonder1982}).
Nevertheless, the expressions for coefficients in the strong-barrier case
are the same for the SM-EI and the N-S junction.

The appearance of coherence factors $u$ and $v$ in the coefficients 
of the SM-EI junction demonstrates that the electron-hole condensate strongly
affects the transport and in general the wave function of carriers,
by means of both inducing coherence on the SM side and altering
transmission features.

\subsection{Proximity effect}\label{s:proximity}

From the results of the previous two subsections it follows that
the condensate on the right hand side of the junction induces
pairing correlations in the normal phase on the left hand side,
even if there the order parameter $\Delta$ is zero as interactions
are switched off. In the N-S junction the pairing 
induced by the proximity effect
correlates electrons with opposite spins, whereas in the SM-EI
junction it correlates electron-hole pairs.

To show that the exciton condensate induces exciton order
on the SM side it is sufficient to 
compute the 
built-in coherence
\protect{$\langle\Psi_0|\tilde{\psi}_b\!\left(\bm{r} \right)
\tilde{\psi}_a^{\dagger}\!\left(\bm{r}\right)
\!|\Psi_0\rangle$}. This would be zero in
an isolated SM but by the proximity with the condensate
acquires the value
\begin{eqnarray}
&& \langle\Psi_0|\tilde{\psi}_b\!\left(\bm{r} \right)
\tilde{\psi}_a^{\dagger}\!\left(\bm{r}\right)
\!|\Psi_0\rangle  =  \sum_{\bm{k}}f_{\bm{k}}\!\left(\bm{r}\right)
g^*_{\bm{k}}\!\left(\bm{r}\right) \nonumber\\
& \approx & 2\int\!\!\text{d}\,\omega\,N\!\left(\omega\right)
\cos{\left[\arctan{\left(\frac{\Delta}{\omega}\right)}+2
\frac{\omega}{v_{\text{F}}}z\right]},
\label{eq:sum}
\end{eqnarray}
with $N\!\left(\omega\right)$ being the density of states
in the SM.
Inside the gap ($\omega\approx 0$) each quasiparticle
contributes to the sum (\ref{eq:sum})
with a term $\sim \exp{\left[\text{i}\arctan{\left(\Delta/\omega\right)}
+2\text{i}\omega z/v_{\text{F}}\right]}$. The only coordinate dependence
enters this expression via the phase factor, $2\omega z/v_{\text{F}}$,
which represents the relative phase shift of conduction- and
valence-band components of the wave function.
If $\omega=0$, then these components keep constant relative phase
$\arctan{\left(\Delta/\omega\right)}$
all the way to $z=-\infty$, where no pairing interactions
exist.
Therefore, the reflected electron---equivalent to an incoming hole---has 
exactly the
same velocity as the incident particle,
and will thus retrace {\em exactly the same path} all the way
to $z=-\infty$. At finite energy,
the $z$ dependent oscillations provide destructive interference
on the pair coherence.
Hence, the paths of incident and reflected electrons
part ways away from the interface. Analogous considerations apply to
the incident electron and to the Andreev-reflected hole in a sub-gap
scattering event at the N-S interface \cite{Zagoskin1998}.

\section{A perfect insulator}\label{s:results}

From the
results for transmission and
reflection probabilities obtained in Sec.~\ref{s:transmission},
we derive in the linear response regime the values of the electrical and
thermal conductances of the SM-EI junction, $G$ and $G_T$,
respectively. The derivation is standard and it proceeds along the lines
explained e.g.~in Refs.~\cite{Blonder1982,Datta1995,Rontani2009b}. 
The Seebeck coefficient is zero due to the symmetry of the model 
\cite{Ziman1960,Rontani2009b}.
Then, except for an additive phonon contribution to the thermal conductance,
the interface thermoelectric properties are completely determined by
$G$ and $G_T$.

\begin{figure}[tb]
\setlength{\unitlength}{1 cm}
\begin{picture}(8.5,6.5)
\put(1.9,0.0){\epsfig{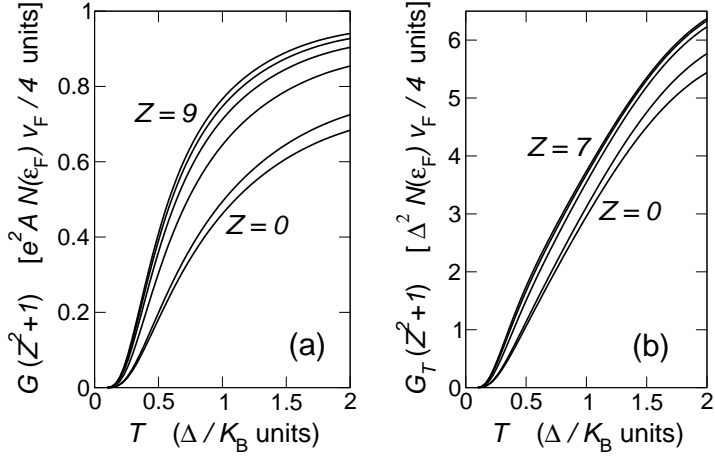}}
%\put(1.8,1.6){(a)}
%\put(8.0,1.6){(c)}
%\put(5.0,1.6){(b)}
\end{picture}
%\centerline{\epsfig{file=./figure3.eps,width=3.1in,,angle=0}}
\caption{
(a) Electrical conductance $G$ of the SM-EI junction
vs temperature $T$. The curves correspond to different
values of the dimensionless barrier opacity,
\protect{$Z=0,1,3,5,7,9$}.
For each value of $Z$, $G$ is
divided by the corresponding transmission coefficient
when $\Delta=0$, i.e.,
$C_{\Delta=0}=(Z^2+1)^{-1}$.
$K_{\text{B}}$ is the Boltzmann constant, $\mathcal{A}$ is the interface
cross-sectional area, $N(\varepsilon_{\text{F}})$ is the density of 
states evaluated at the Fermi energy $\varepsilon_{\text{F}}$.
(b) Thermal conductance $G_T$ of the SM-EI junction
vs $T$. 
The curves correspond to \protect{$Z=0,1,3,5,7$}.
}
\label{figGT}
\end{figure}

Both $G$ and $G_T$ have an activation threshold at low temperature,
$T$, proportional
to the gap $\Delta$, as shown in Fig.~\ref{figGT} 
(cf.~the curves for the transparent barrier with $Z=0$).
At first sight, the transport properties of the SM-EI junction
seem dramatically different from those of the N-S junction,
as the latter may sustain an electric supercurrent at
vanishing bias voltage whereas the former 
exhibits insulating behavior. However, a closer examination
shows that the two junctions share essential features.  

In fact,
the functional dependence of $G_T$
on $T$ and $\Delta$ is the same 
for both the SM-EI and N-S junctions.
Remarkably, the time rate of entropy production, $\dot{S}$, 
is the same very low value in both cases, pointing to the
disipationless character of the flow of charge and heat. 
This is seen by the relation between $\dot{S}$ and the transport
coefficients \cite{Ziman1960},
\begin{equation}
\dot{S} = G (\delta V)^2/T + {\mathcal A} G_T (\delta T)^2/T^2,
\label{eq:Sdot}
\end{equation}
with $\delta T$ and $\delta V$ being
respectively  the temperature and voltage drops
at the interface and ${\mathcal A}$ being the cross-sectional area.
In the N-S junction the superfluid component does not carry any entropy.
Therefore, the terms proportional to $G$ and $G_T$ 
only include the contribution of quasiparticles which,
when they cross the N-S interface, experience
the same electric and thermal resistance as 
electrons do across the SM-EI boundary.

\begin{figure}[tb]
\setlength{\unitlength}{1 cm}
\begin{picture}(8.5,7.5)
\put(1.9,0.5){\epsfig{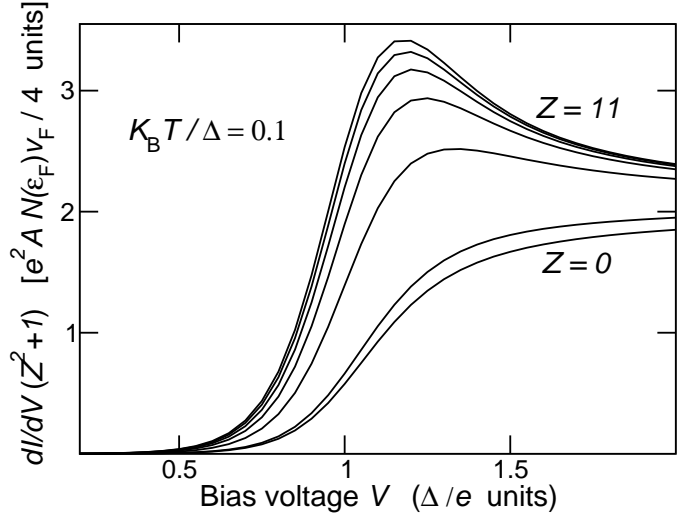}}
%\put(1.8,1.6){(a)}
%\put(8.0,1.6){(c)}
%\put(5.0,1.6){(b)}
\end{picture}
%\centerline{\epsfig{file=./diffcond.eps,width=3.1in,,angle=0}}
\caption{
Differential conductance $(dI/dV)/C_{\Delta=0}$ of the SM-EI junction
vs bias voltage $V$, computed at $K_{\text{B}}T/\Delta = 0.1$. 
The curves correspond to different
values of the dimensionless barrier opacity,
\protect{$Z=0,1,3,5,7,9,11$}.
For each value of $Z$, $dI/dV$ is
divided by the corresponding transmission coefficient
when $\Delta=0$, i.e.,
$C_{\Delta=0}=(Z^2+1)^{-1}$.
}
\label{diffcond}
\end{figure}

To shed light on the dissipationless motion of electrons
in the linear transport regime,
we vary the opacity of the SM-EI junction.
The coherence between the two sides of the interface
is diminished as the dimensionless barrier strength
$Z$ increases from zero (clean junction)
to finite values (tunneling regime).
Figure \ref{figGT} displays $G$ and $G_T$
for increasing values of $Z$. Since the transmission coefficient $C(\omega)$
decreases uniformly in the absence
of any electron-hole pairing (cf.~Table \ref{tab:table2}),
$C_{\Delta=0}=(Z^2+1)^{-1}$,
we rescale conductances dividing them by $C_{\Delta=0}$.
Naively, we would expect that the insertion of an insulating layer
would reduce the conductances.
On the contrary, the effect is just the opposite: as $Z$ increases,
$G/C_{\Delta=0}$ and $G_T/C_{\Delta=0}$ {\em increase}, eventually
reaching saturation in the tunneling regime.
This shows that the exciton order induced in the SM side by EI
makes the junction less conductive for charge and heat transport.
The plot of the differential conductance $(dI/dV)/C_{\Delta=0}$
vs the bias voltage $V$ at low $T$ (Fig.~\ref{diffcond})
allows clear monitoring of the transition from the transparent to the
opaque limit, where transport is recovered.
The effect is maximum for $eV\approx \Delta$ and as $T\rightarrow 0$,
when the differential conductance becomes proportional to
$C(eV)+D(eV)$.

\subsection{Charge versus exciton current}

As anticipated at the end of Sec.~\ref{s:velocity},
the transport features discussed above 
which distinguish the excitonic insulator from the normal
insulating state may be explained by two alternate physical pictures.
The conventional view is that electrons below the energy gap cannot
contribute to transport as they are back scattered by the gap
barrier, $\Delta$, formed by the proximity effect of the
EI. The less conventional view is to make use of the
analogy with the N-S junction. Instead of
counting the electrons in the valence band as negatively charged
carriers of the current, we may start with the state with the valence
band filled to the top as carrying zero current even under an
electrical or thermal current and regard each unoccupied state in the
valence band as a positively charged carrier --- a hole --- moving in the
direction opposite to the electron. Then the reflected electrons are
replaced by incoming holes towards the barrier.
Therefore, the incident conduction electron and the valence hole may
be viewed as a correlated pair moving  towards the interface
[Fig.~\ref{equivalence}(a)].
The novelty is that a constant
electron-hole current moves from the SM to the EI below the gap, where
electric transport is blocked.
As the electron-hole pair approaches the interface
from the SM side, the {\em exciton} current
is converted into the condensate {\em supercurrent}:
the global effect is that in the steady state an exciton current exists
flowing constantly and reversibly all the way from the SM to the EI
without any form of dissipation.

The above scenario follows from the continuity equation
for the electron-hole current. The probability density
$\rho_{\text{e-h}}\!\left(\bm{r},t\right)$ for finding either
a conduction-band
electron or a valence-band hole at a particular time and place is
\protect{$\rho_{\text{e-h}}\!\left(\bm{r},t\right)=
\left|f\right|^2 + 1 - \left|g\right|^2$}.
Thus, the associated continuity equation is
\begin{equation}
\frac{\partial \rho_{\text{e-h}}}{\partial t}+\nabla\cdot
{\bm{J}_{\text{e-h}}}=0, \qquad
\bm{J}_{\text{e-h}} = \bm{J}_{\text{pair}} + \bm{J}_{\text{cond}}.
\label{eq:continuity2}
\end{equation}
One component of the electron-hole current,
\begin{equation}
\bm{J}_{\text{pair}}=\frac{1}{m}{\text{Im}}\left\{
f^*\nabla f + g^*\nabla g\right\}, 
\end{equation}
is the
density current of the electron-hole pair, similar to
the standard particle carrier \protect{$\bm{J}=m^{-1}{\text{Im}}\{
f^*\nabla f - g^*\nabla g\}$} with an important difference in sign.
The other component,  
\begin{equation}
\nabla\cdot {\bm{J}_{\text{cond}}} =
-4\,{\text{Im}}\left\{ f^* g \, \Delta \right\},
\end{equation}
depends explicitly on the built-in coherence of the electron-hole
condensate $\Delta$, and may be described as the exciton
supercurrent of the EI state.

Let us go back to our picture of $\Delta\!\left(z\right)$ smoothly varying
in space (Fig.~\ref{figura2}), with the junction being divided
into small neighborhoods at position $\bm{r}$ and each being a
homogeneous system.
If $\omega < \left|\Delta_0\right|$,
each electron wave function, solution of eqn (\ref{eq:BdG2}),
carries zero total electric current $e\bm{J}$, which is the
sum of the equal and opposite
incident and reflected fluxes, and finite and constant
electron-hole current \protect{$\bm{J}_{\text{e-h}}=2v_{\text{F}}\bm{n}$},
with $\bm{n}$ a unit vector.
When $z\rightarrow -\infty$, far from the interface on the SM side,
the supercurrent contribution $\bm{J}_{\text{cond}}$
is zero.
As $z$ increases and $\Delta\!\left(z\right)$ gradually rises,
both $\bm{J}$ and \protect{$\bm{J}_{\text{e-h}}$} conserve
their constant value, independent of $z$, since quasiparticle states
are stationary. However, their analysis in terms of incident and
reflected quasi-particles is qualitatively different.

From the electron point of view, we see in Fig.~\ref{figura2}(a) that
the incoming conduction-band particle approaching the EI boundary
sees its group velocity progressively reduced (from time step 1
to time step 3), up to the classical
turning point (time step 4)
where it changes direction and branch of the spectrum:
there is no net electric current.

From the exciton point of view,
as the contribution to the electron-hole current $\bm{J}_{\text{pair}}$
vanishes approaching the boundary, since the group velocity
goes to zero at the classical turning point
where the wave function becomes evanescent, $\bm{J}_{\text{pair}}$ is
converted into the supercurrent $\bm{J}_{\text{cond}}$.
Excitons therefore can flow into the EI side without any resistance,
and the sum \protect{$\bm{J}_{\text{e-h}}$} of the two contributions,
$\bm{J}_{\text{pair}}$ and $\bm{J}_{\text{cond}}$,
is constant through all the space [Fig.~\ref{equivalence}(a)].

As $\omega$ exceeds $\left|\Delta_0\right|$, $\bm{J}$ acquires a finite
value and \protect{$\bm{J}_{\text{e-h}}$} monotonously decreases.
However,
close to the gap, electron transmission to the EI side is
still inhibited [cf.~eqn (\ref{eq:Cdep})] by the pairing between
electrons and holes of the condensate: an electron can stand alone and
carry current only after its parent exciton has been ``ionized''
by injecting --- say ---  a conduction-band electron or by filling
a valence-band hole in the EI. The ionization costs
an amount of energy of the order of the binding energy of the
exciton, $\left|\Delta_0\right|$.
Therefore, as long as $\omega \approx \left|\Delta_0\right|$,
the competition between exciton and electron flow favors
Andreev
reflection, which is the source of both electric and thermal resistances.

In equilibrium, there is no net charge or heat
flow, since quasi-particles with $\bm{v}$ and $-\bm{v}$
compensate each other. However, if a heat current flows,
the net drift velocity of electrons and holes locally
``drags'' the exciton supercurrent, which otherwise would be pinned
by various scattering sources \cite{Guseinov1973}.

\subsection{A concrete example}

As a concrete example of the aforementioned conversion 
of free-exciton current into condensate supercurrent,
consider the quasiparticle steady
state of Eq.~(\ref{eq:boundary}).
For $\omega<\Delta$, $k^+$ and $k^-$ in the EI
($z>0$) have small imaginary parts which lead to an exponential decay on
a length scale $\lambda$, where
\begin{equation}
\lambda = \frac{v_{\text{F}}}{2\Delta}\left(1-\frac{\omega^2}{\Delta^2}
\right)^{-1/2}.
\end{equation}
The quasiparticles penetrate a depth $\lambda$ before
the electron-hole current $J_{\text{pair}}$ is converted to
a supercurrent $J_{\text{cond}}$ carried by the condensate; right
at the gap edge the length diverges. For clarity, we define
$C$ and $D$ here as the transmission probabilities at $z\gg \lambda$,
while for $\omega>\Delta$ plane-wave currents are spatially uniform
and we need not specify the position at which they are evaluated.

When the interface is transparent, $Z=0$, the steady state
(\ref{eq:boundary}) is specified by $b=d=0$, $a=v_0/u_0$,
and $c=1/u_0$. Below the gap
coherence factors $u_0$ and $v_0$ are complex and
equal in modulus. For $\omega<\Delta$, $\left|a\right|^2=1$, which
means the incident conduction-band electron is totally
reflected into the SM valence band. Thus, the electron-hole
current  $J_{\text{pair}}$ carried in the semimetal equals
$2v_{\text{F}}$, but $J_{\text{pair}}$ of the excitonic insulator
is exponentially small for $z\gg 0$. Explicitly,
\begin{displaymath}
J_{\text{pair}} = \frac{\left|c\right|^2}{m}(\left|u_0\right|^2+
\left|v_0\right|^2)\,{\text{Im}}\!\left[({\rm e}^{ \text{i} k^+z }
)^*\frac{\partial}{\partial z} ({\rm e}^{ \text{i} k^+z })\right].
\end{displaymath}
Letting \protect{$k^+\approx k_{\text{F}} + {\text{i}}/(2\lambda)$},
we have
\begin{equation}
J_{\text{pair}} = 2v_{\text{F}}{\rm e}^{-z/\lambda}.
\end{equation}
The ``disappearing'' electron-hole current reappears as
exciton current carried by the condensate. Recalling the
definition of $J_{\text{cond}}$,
\begin{displaymath}
\partial J_{\text{cond}} / \partial z =
-4\,{\text{Im}}\!\left\{ f^* g \, \Delta \right\},
\end{displaymath}
by integration we obtain
\begin{equation}
{J}_{\text{cond}} = -4\Delta\left|c\right|^2\int_0^z\!\!{\text{d}}\,z'
{\rm e}^{-z'/\lambda}\,{\text{Im}}\!\left[u_0^*v_0\right]
= 2v_{\text{F}}\left(1-{\rm e}^{-z/\lambda}\right).
\end{equation}
This is the desired result, explicitly showing the supercurrent
${J}_{\text{cond}}$
increasing to an asymptotic value as $z\rightarrow \infty$, at the
same rate as the free electron-hole current $J_{\text{pair}}$ dies away.

\section{Josephson oscillations between exciton condensates
in electrostatic traps}\label{s:Josephson} 

The Josephson effect is a a macroscopic coherent phenomenon which has been 
observed in systems as diverse as superconductors \cite{Barone1982},
superfluid Helium \cite{Avenel1985,Pereverzev1997,Sukhatme2001},
and Bose-Einstein condensates of 
trapped ultra-cold atomic 
gases \cite{Cataliotti2001,Shin2004,Albiez2005,Levy2007}.
Since Josephson oscillations appear naturally when 
two spatially separated macroscopic wave functions are weakly coupled,
they have been predicted for  bosonic excitations in solids as well, like 
polaritons \cite{Carusotto2007,Sarchi2008} and excitons optically
pumped in suitable semiconductor heterostructures \cite{Shelykh2008}.
However, unlike the polaritons, which have a photonic component 
allowing for easy detection, excitons stay dark unless they
recombine radiatively. So far, it is unclear how the exciton
Josephson effect could be observed. One proposal relied on
the observation of quasiparticle excitations from the spectral properties 
of the emitted light \cite{Sarchi2008}. 
A drawback of this idea is that spectral
properties are not unambiguously linked to the condensed phase.
In this Section we propose the interference of
emitted photons as a direct probe of the exciton Josephson effect. 

Condensed excitons are predicted to act as coherent 
light sources \cite{Oestreich1996,Rossier1998,Castro2001}
(see discussion below). 
If Josephson oscillations occur
between two exciton traps, in principle they can be probed by 
measuring the interference of the beams separately emitted from the traps. 
However, in the time interval before recombination and contrary to
the polariton case, 
there are too few photons emitted for an adequate signal to noise ratio,
so one has to average the signal over many replicas of the same 
experiment \cite{High2012}.
Here we show that such ensemble averaging
blurs the signature of the Josephson effect 
except in the relevant case of exciton ``plasma'' 
oscillations \cite{Stringari2003}. For the latter the
dipole energy difference between the traps modulates 
the visibility $\alpha$
of interference fringes, providing a means for detection. 

The subsections below are organized as follows:
we first introduce the double quantum well system
illustrating a feasible scheme to manipulate electrically
the exciton phase (Sec.~\ref{s:electrical}); we then 
set the theoretical framework 
(Sec.~\ref{s:JosephsonTh}), and we 
eventually propose a correlated photon
counting experiment (Sec.~\ref{s:correlated}).

\subsection{Electrical control of the exciton phase}\label{s:electrical}

\begin{figure}
\setlength{\unitlength}{1 cm}
\begin{picture}(8.5,5)
\put(0.2,0.2){\epsfig{file=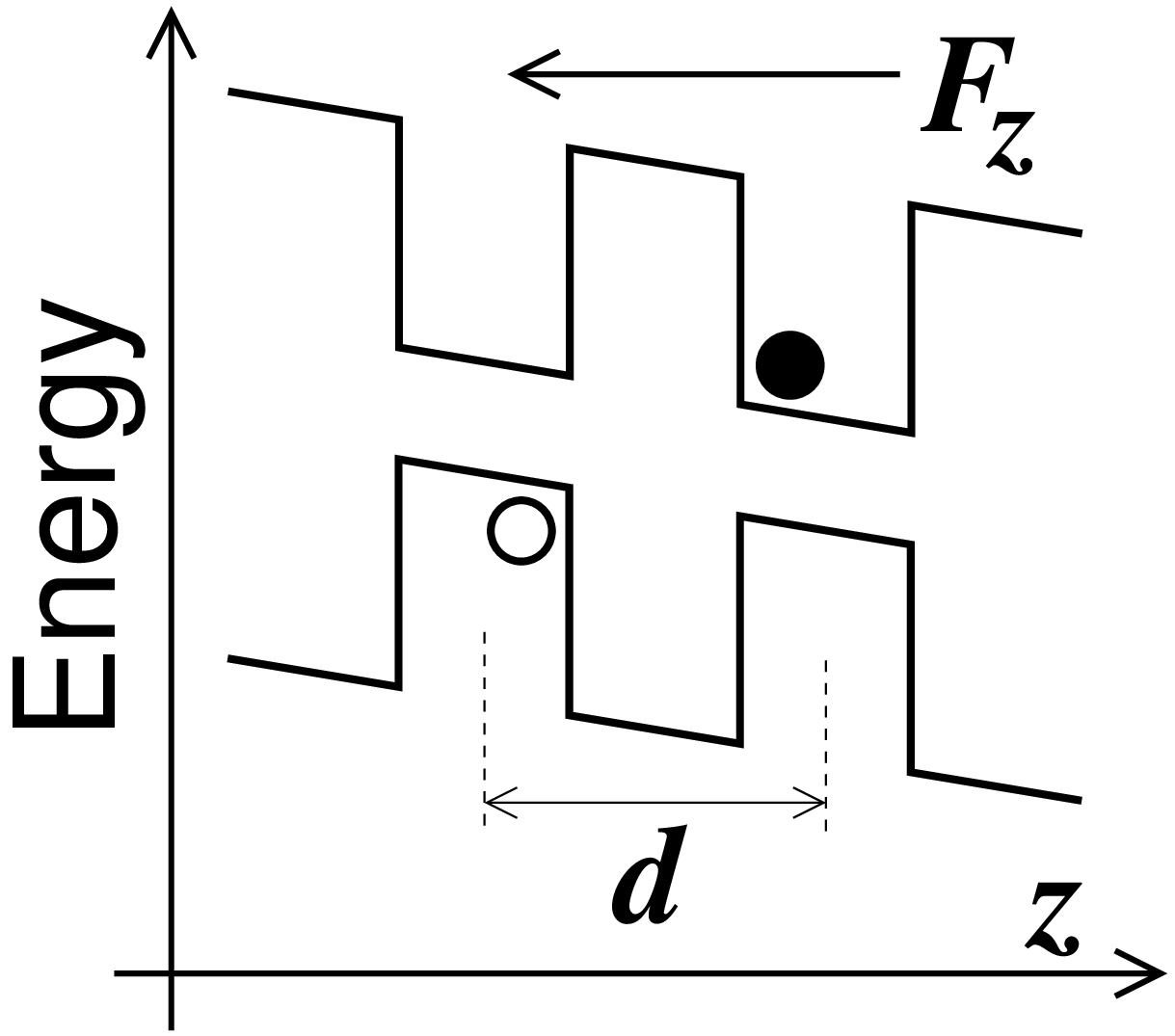,angle=-0,width=4.2cm}}
\put(6.2,0.2){\epsfig{file=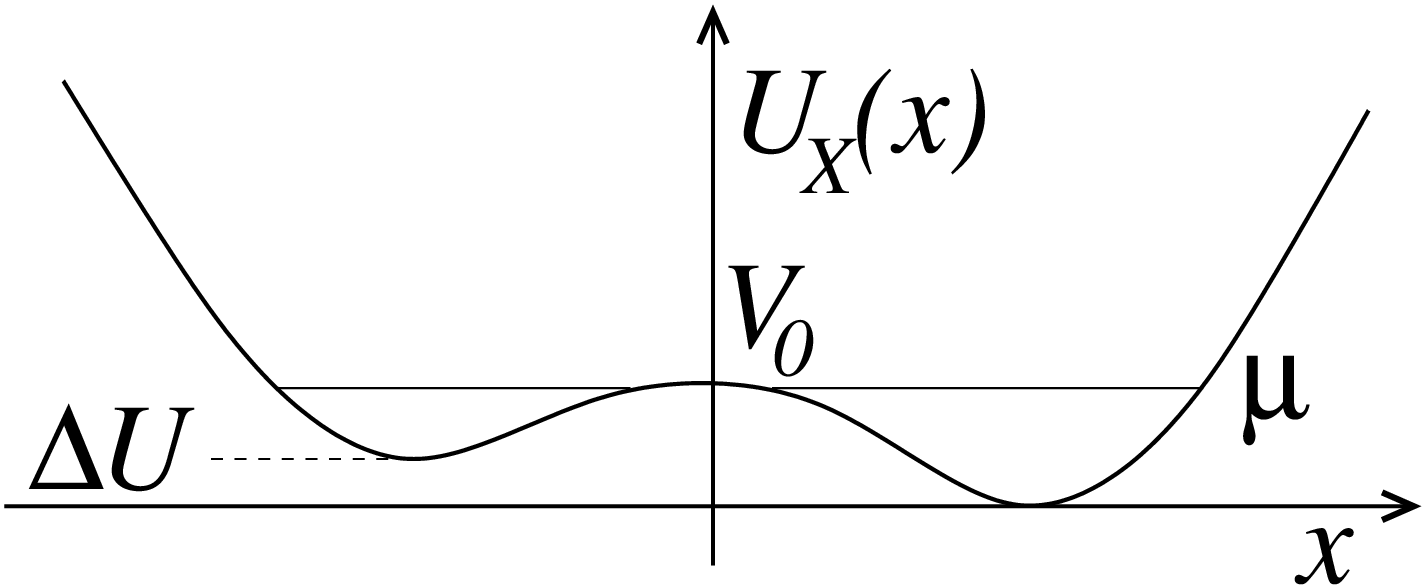,angle=-0,width=6.5cm}}
%\put(5.7,0.0){\epsfig{file=fig1c.eps,angle=-0,width=2.5cm}}
\put(4.8,3.6){(a)}
%\put(8.0,1.6){(c)}
\put(12.6,3.6){(b)}
\end{picture}
\caption{(a) Double quantum well energy profile along the growth
direction $z$. Electrons and hole move freely in the $xy$ planes
at $z=d$ and $z=0$, respectively. $F_z$ is the component of the
electric field parallel to the exciton electric dipole.
(b) Effective exciton potential profile vs $x$ for the 
double-trap setup.
}
\label{fig1}
\end{figure}

Consider a double quantum well where electrons and holes 
are separately confined in the two layers. In experiments
aiming at Bose-Einstein condensation of excitons, 
electron-hole pairs are optically generated at energies
higher than the band gap, 
left to thermalize, form excitons, 
and, at sufficiently low temperature and high density,
condense before radiative decay 
\cite{Ivanov2008,High2012}.
Let $z$ be the growth axis of the two wells separated by 
distance $d$. The electrons in the conduction band and 
holes in the valence band move in the planes $z=d$, $0$, 
respectively (Fig.~\ref{fig1}). 
In the experiments \cite{Butov2004,Butov2007},
an electric field $F_z$ is applied along $z$ 
to suppress inter-layer tunneling, thereby 
quenching the exciton recombination.

Fabrications \cite{Hammack2006,High2008,Chen2006} 
of electrostatic traps with suitably located 
electrodes to provide lateral confinement for the 
excitons have been implemented. 
The double quantum well is sandwiched between two spacer layers, 
providing insulation from planar electrodes lithographed on both sides of 
the coupled structure.
Each electrode controls a tunable gate voltage, $V_g(x,y,z)$, which 
localizes in a region of the $xy$ plane the field component 
along $z$, $F_z(x,y,z)=-\partial V_g(x,y,z) /\partial z$, 
while $F_x$ and $F_y$ are 
small and can be neglected as well as the the dependence of $F_z$ on $z$. 
The vertical field $F_z(x,y)$  makes the electrostatic potential energy 
of the exciton dipole 
depend on the lateral position, $U_X(x,y)=-edF_z(x,y)$ ($e<0$)
[cf.~Fig.~\ref{fig1}(b)].
In this way, potential traps for excitons are 
designed with great flexibily, with
{\em in situ} control of the height, width, and shape of the 
potential barriers \cite{Hammack2006,High2012b}.

First, we focus on the quasi-equilibrium situation before radiative 
recombination, where excitons  condense in two coupled electrostatic 
traps, both within the condensate coherence length.
Figure \ref{fig1}(b) depicts the exciton potential 
profile $U_X(x,y=0)$ along the $x$ axis, with a link between two identical 
traps. The potential barrier allows tunneling between the 
condensates in the two traps, whose macroscopic wave functions 
are $\Xi_1(x,y,t)$ and $\Xi_2(x,y,t)$, respectively.
The optical coherence in a single trap is of the form
\begin{equation}
\Xi(x,y,t) = \left< \Psi^{\dagger}_a(x,y,0,t) \Psi_b(x,y,d,t) \right>\;,
\label{eq:definition}
\end{equation}
where $\left<\ldots\right>$ denotes quantum and thermal average
in the grand canonical formalism. 
In eqn (\ref{eq:definition}) $\Psi^{\dagger}_a(x,y,0,t)$ and 
$\Psi_b(x,y,d,t)$ are respectively the hole and
electron destruction operators, with the vacuum being the 
semiconductor ground state with no excitons. 
With respect to definition (\ref{eq:delta}), here by putting 
$(x,y)=(x^{\prime},y^{\prime})$ we ignore
the internal structure of the exciton relative-motion wave function.
The reason is that we focus on the ideal BEC limit of 
dilute weakly interacting excitons,
$n\, a_{\text{B}}^2
\ll 1$, with $a_{\text{B}}$ being the two-dimensional
effective Bohr radius and $n$ the exciton density. 
Therefore, $\Xi(x,y,t)$ is the macroscopic wave function
for the center-of-mass motion of excitons,
which may be written in the form
\begin{equation}
\Xi(x,y,t) = \sqrt{n_s}\; e^{{\rm i}\varphi},
\end{equation}
with $n_s$ being the density of the exciton condensate and $\varphi$ the
phase. 
We recall that only the relative phase between the two condensates 
has measurable effects \cite{Anderson1966,Leggett1991}. 

For a gauge transformation of the gate potential 
$V_g\rightarrow V_g -c^{-1}\partial \chi(t)/\partial t $,
which leaves the field $F_z$ unaltered, 
the field operators $\Psi$ gain a phase,
\begin{eqnarray}
\Psi_{a} &\rightarrow & 
\Psi_{a} \,\exp{\!\left[\!\frac{{\rm i}e}{\hbar c}\,\chi(x,y,0,t)\right]}\;, 
\nonumber\\ 
\Psi_{b} & \rightarrow & \Psi_{b} \,\exp{\!\left[\frac{{\rm i}e}{\hbar c}
\,\chi(x,y,d,t)\right]}\; .
\end{eqnarray}
Throughout this section we indicate explicitly the
reduced Planck's constant $\hbar$.
The macroscopic wave function, by eqn (\ref{eq:definition}), 
also gains a phase,
\begin{equation}
\varphi \rightarrow \varphi + \frac{e}{\hbar c}\left[\chi(z=d,t)
-\chi(z=0,t)\right]\;.
\label{eq:phitransf}
\end{equation}
Hence, the frequency of time oscillation of the condensate 
is given by the electrostatic energy of the exciton dipole
in the external field \cite{Balatsky2004}, $U=-edF_z$:
\begin{equation}
\varphi = \varphi^{(0)} + \frac{1}{\hbar}edF_zt\;,
\label{eq:phitime}
\end{equation} 
with $\varphi^{(0)}$ being 
the time-independent zero-field value.
In the absence of the bilayer separation of the electrons 
and the holes, their gauge phases 
gained in the electric field would cancel each other
resulting in no time dependence driven by $U$.
Equation (\ref{eq:phitime}) shows that the experimentally controllable 
dipole energy difference between the two traps depicted 
in Fig.~\ref{fig1}(b),  $\Delta U= -ed(F_{z1}-F_{z2})$, 
drives the relative phase between the two condensates, 
thereby creating Josephson oscillations as a means for 
measuring the Josephson tunnel between the traps. 

\subsection{Exciton Josephson oscillations}\label{s:JosephsonTh}

We next introduce the usual two-mode description of inter-trap
dynamics based on the Gross-Pitaevskii (GP)
equation 
\cite{Carusotto2007,Sarchi2008,Shelykh2008,Smerzi1997,Raghavan1999,Zapata1998,Stringari2003}.
Exciton-exciton correlation 
\cite{Oestreich1999} beyond the GP mean field  
may be neglected due to the repulsive character of
the dipolar interaction 
between excitons in coupled quantum wells.
The condensate total wave function solution is
\begin{equation}
\Xi(x,y,t) = \Xi_1(x,y,N_1)\,\text{e}^{i\varphi_1} +
\Xi_2(x,y,N_2)\,\text{e}^{i\varphi_2}\;,
\label{eq:two}
\end{equation}
where both the trap population $N_i(t)$ and the condensate phase 
$\varphi_i(t)$ 
possess the entire time dependence for the $i$th trap  ($i=1,2$),  and 
$\Xi_i(x,y,N_i)$ is a real quantity, with 
\begin{equation}
\int\!\! dx\!\!\int\!\! dy\; \Xi_i^2(x,y,N_i) = N_i(t).
\end{equation} 
The dynamics
of the GP macroscopic wave function
$\Xi(x,y,t)$ depends entirely on
the temporal evolution of two variables, the population imbalance
$k(t)=(N_1-N_2)/2$ and the relative phase $\phi(t)=\varphi_1-\varphi_2$
of the two condensates. Here we  consider a
time interval much shorter than the exciton lifetime
%%($\sim 10 \div 100$ ns) 
(10 --- 100 ns) and ignore the spin structure.
Therefore, the total population is approximately constant,
$N_1(t)+N_2(t)=N$. 

The equations of motion for the
canonically coniugated variables $\hbar k$ and $\phi$
are derived from the effective Hamiltonian
\begin{equation}
H_J = E_c\frac{k^2}{2} +\Delta U k
-\frac{\delta_J}{2}\sqrt{N^2-4k^2}\cos{\phi}\;,
\label{H}
\end{equation}
under the condition $k\ll N$ (Ref.~\cite{Stringari2003}).
$E_c=2d\mu_1/dN_1$ is the exciton ``charging'' energy of one trap, 
where $\mu_1$ is the chemical potential of trap 1,
whereas $\delta_J$ is the Bardeen single-particle tunnelling energy,
\begin{equation}
\delta_J=\frac{\hbar^2}{m}\int\!\!dy\left[\xi_1\!
\left(\frac{\partial \xi_2}{\partial x}\right)
-\xi_2\!\left(\frac{\partial \xi_1}{\partial x}\right)\right]_{x=0},
\end{equation}
where $m$ is the exciton mass. The single-particle orbital $\xi_i(x,y)$
is defined through $\Xi_i(x,y)=\sqrt{N_i}\,\xi_i(x,y)$.

The various dynamical regimes associated to certain
intitial conditions $\left(k(0),\phi(0)\right)$,
including $\pi$ oscillations
and macroscopic quantum self-trapping, are 
exhaustively discussed in Refs.~\cite{Smerzi1997,Raghavan1999}. 
Two cases are specially relevant:

\subsubsection{AC Josephson effect} 

Under the conditions $\Delta U \gg NE_c/2$,
$\Delta U\gg \delta_J$, one easily obtains
\begin{equation}
\phi(t) = -\frac{\Delta U}{\hbar}t+\phi(0)\;,\qquad 
\dot{k} =\frac{\delta_JN}{2\hbar}
\sin{\phi}\;.
\label{AC}
\end{equation}
Equation (\ref{AC}) shows that, analogous to the 
case of two superconductors separated by a thin barrier, if the
phase difference $\phi$ between the condensates is not a multiple
of $\pi$, an exciton supercurrent $2\dot{k}$ flows across the barrier. 
Remarkably, in the presence of an electric field gradient along $z$,
an exciton flux oscillates back and forth between the two traps,
with frequency $\Delta U/\hbar$. 
As an exciton goes through the barrier, it exchanges with
the field the dipole energy acquired or lost in the 
tunneling process.
The analogy with the AC Josephson effect for
superconductors is clear: in that case 
a bias voltage $V$ is applied across the junction,
and the energy $2eV$ is exchanged between field and 
Cooper pairs, as the latter experience a potential
difference of $V$ when penetrating the potential barrier.

\subsubsection{Plasma oscillations}

This case concerns small oscillations around
the equilibrium position $(k,\phi)_{\text{eq}}=(0,0)$. 
The Hamiltonian (\ref{H}) may then be linearized into the form
\begin{equation}
H_J = \frac{k^2}{2}\left(2\frac{\delta_J}{N}+E_c\right)+\frac{1}{4}\delta_J
N\phi^2+\Delta U k-\frac{\delta_JN}{2}\;.
\label{Hnormal}
\end{equation}
It follows that both $k$ and $\phi$ oscillate in time 
with plasma frequency 
\begin{equation}
\omega_J=\frac{1}{\hbar}\sqrt{\delta_J(NE_c/2+\delta_J)}. 
\end{equation}
Note that $\Delta U$ displaces the equilibrium position
from $(k,\phi)_{\text{eq}}=(0,0)$ to 
\begin{equation}
(k,\phi)_{\text{eq}}=(-\Delta U N \delta_J/2(\hbar\omega_J)^2,0).  
\label{eq:displace}
\end{equation}

\subsection{Correlated photon counting experiment}\label{s:correlated}

\begin{figure}
\setlength{\unitlength}{1 cm}
\begin{picture}(8.5,5)
\put(4.2,0.7){\epsfig{file=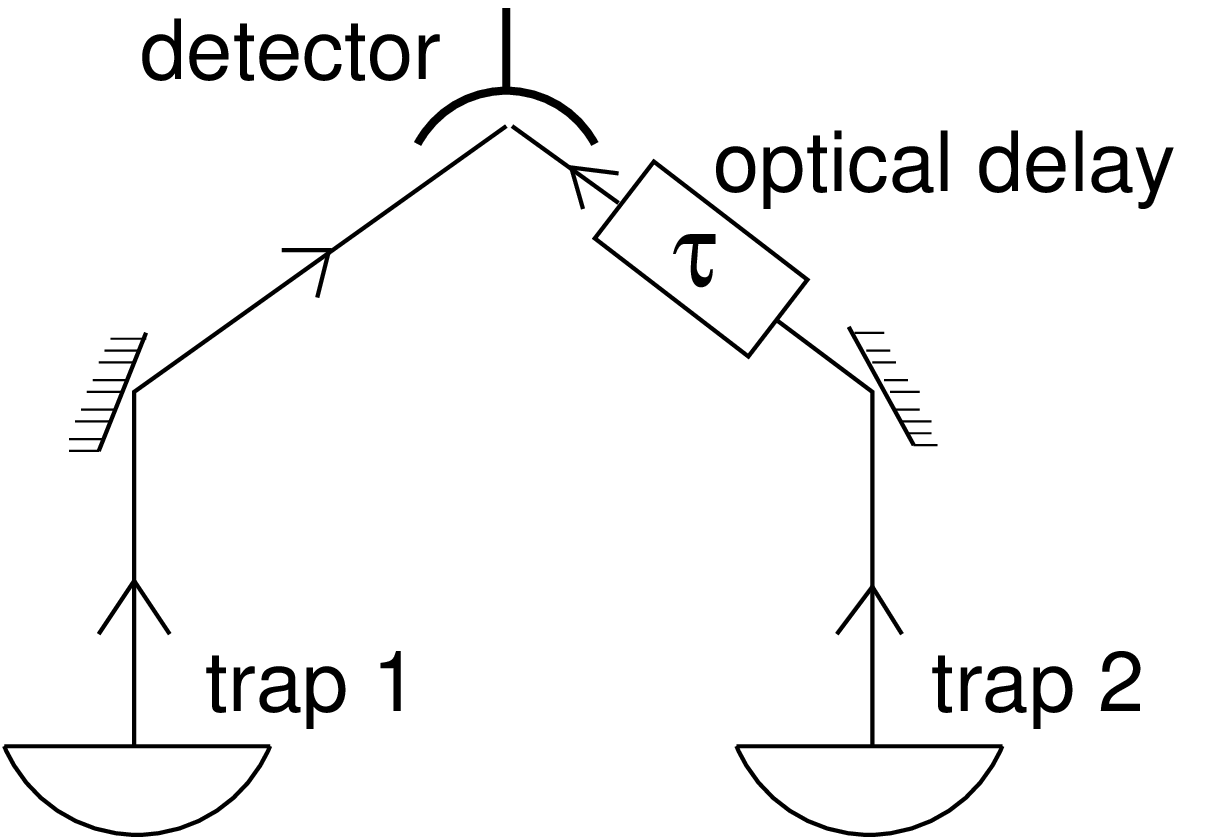,angle=-0,width=5.2cm}}
%\put(6.2,0.2){\epsfig{file=fig1b.eps,angle=-0,width=6.5cm}}
%\put(5.7,0.0){\epsfig{file=fig1c.eps,angle=-0,width=2.5cm}}
\put(4.4,0.3){$\Xi_1,\varphi_1$}
\put(7.5,0.3){$\Xi_2,\varphi_2$}
%\put(8.0,1.6){(c)}
%\put(12.6,3.6){(b)}
\end{picture}
\caption{Proposed experimental setup to measure
the time correlation of the photons emitted from the two traps.
The delay time $\tau$ of one optical path is
externally controlled.
}
\label{setup}
\end{figure}
Figure \ref{setup} illustrates 
the correlated photon counting setup which
we propose to probe Josephson oscillations. The detector
measures the intensity $I(\tau)$ 
of the sum of the two beams separately emitted from
the traps. A delay time $\tau$ is induced in one of the two
beams, as in Ref.~\cite{Yang2006}. 
The fields are simply proportional
to the order parameters $\Xi_i$ of the traps. In fact, $\Xi(x,y,t)$
is associated with a macroscopic
electric dipole
moment, 
\begin{equation}
\bm{P}(t)=\hat{\bm{x}}P_x(t)\pm i\hat{\bm{y}}P_y(t),
\end{equation}
which couples to photons:
\begin{equation}
P_{x}(t)=\int dx\,dy \, x\,\, \Xi(x,y,t),
\end{equation}
and similarly for $P_y$.
The built-in dipole 
$\left<\bm{P}(t)\right>\neq 0$ oscillates
with frequency $(\mu+E_X)/\hbar$,
where $E_X$ is the optical gap minus the exciton binding energy,
and $\mu$ accounts for exciton-exciton interaction
\cite{Oestreich1996,Rossier1998,Castro2001}.
This macroscopic oscillating 
dipole is equivalent to a noiseless current, which 
radiates a coherent field \cite{Glauber1965}.

Therefore, the measured intensity $I(\tau)$ is
\begin{equation}
I(\tau)=2I_0\left[1+\left<\cos{\phi(\tau)}\right> \right],
\end{equation}
assuming that the fields emitted from the two traps have 
the same magnitude (and intensity $I_0$) 
but different relative phase $\phi$, which is evaluated 
at the delayed time $\tau$. $I(\tau)$ may be written as
\begin{equation}
I(\tau)=2I_0\left[1+\alpha\cos{\phi_0(\tau)} \right]\;,
\label{interfe2}
\end{equation}
where $\phi_0(\tau)$ is the phase averaged over many measurements, defined
by the condition 
\begin{equation}
\left<\sin{[\phi(\tau)-\phi_0(\tau)]}\right>=0,
\end{equation}
and 
\begin{equation}
\alpha=\left<\cos{[\phi(\tau)-\phi_0(\tau)]}\right>
\end{equation}
is the fringe visibility,   
i.e., the normalized peak-to-valley ratio of fringes,
\begin{equation}
\alpha=\frac{I_{\text{max}}-I_{\text{min}}}
{I_{\text{max}}+I_{\text{min}}}, 
\end{equation}
with $I_{\text{max}}$ ($I_{\text{min}}$) being
the maximum (minimum) value of $I(\tau)$, and $0\le\alpha\le 1$.

Equation (\ref{interfe2}) has a few important caveats.
Since $I(\tau)$ is an average,
the temporal inhomogeneous effect will 
blur the interference fringes, i.e., $\alpha<1$.
Other dephasing mechanisms include exciton recombination
and inelastic exciton-phonon scattering \cite{Yang2006}, as well as 
inelastic \cite{Yang2006} and elastic
exciton-exciton scattering, 
which in first instance may all be neglected for
short $\tau$, low $T$, and $n\, a_{\text{B}}^2 \ll 1$, respectively. 

The most immediate caveat is that the exciton condensates 
in decoupled traps must acquire a relative phase if initially 
they condense separately without a definite phase relation.
This scenario is analogous to the case of interference
between independent laser sources first discussed by Glauber 
\cite{Glauber1965} and later studied experimentally 
for matter waves \cite{Andrews1997}.
Even though a one-shot measurement with sufficient resolution 
would display fringes, the relative phase $\phi_0(\tau)$ is also 
subject to intrinsic dephasing effects by
quantum fluctuations \cite{Glauber1965}.
The latter are significant noise sources which affect $\alpha$, 
when $\phi$ and $k$ are quantized into canonically
conjugated quantum variables whereas 
in the GP theory used so far they were 
classical variables whose fluctuations where neglected.

In the following, we quantize Hamiltonian (\ref{H})
in order to properly evaluate 
$\alpha=\left<\cos{[\phi(\tau)-\phi_0(\tau)]}\right>$ 
as a quantum statistical average in finite traps.
Therefore, we follow Ref.~\cite{Pitaevskii2001} 
and introduce the commutator
\begin{equation}
\left[\hat{\phi},\hat{k}\right]=i. 
\end{equation}
The operator $\hat{k}$ now appearing in the quantized
version of Hamiltonian (\ref{H}) takes
the form $-i\partial / \partial \phi$,
whereas
the ground state wave function
is defined in the space of periodical functions of $\phi$ 
with period $2\pi$.
If condensate oscillations are mainly coherent, 
the variance of $\phi$ is small and the visibility is 
approximated by $\alpha= 1- \frac{1}{2} \left<(\Delta\phi)^2\right>$. 

The most interesting case concerns plasma oscillations. 
For $\Delta U = 0$, the ground state of the quantized version
of the harmonic
oscillator Hamiltonian (\ref{Hnormal}) is a Gaussian, 
with $\phi_0=0$, independent from $\tau$,
and minimal spreading $\left<\Delta\phi^2\right>\approx 
(E_c/2\delta_J N)^{1/2}$. Therefore, the 
interferometer output is time-independent,
$I=2I_0(1+\alpha)$, showing constructive interference,
$I>2I_0$,
with 
\begin{equation}
\alpha=1-\sqrt{\frac{E_c}{8\delta_J N}}.
\end{equation}
Not surprisingly, the visibility is controlled by
the ratio $E_c/\delta_J N$, reaching the maximum $\alpha =1$
as $ E_c/\delta_J N \rightarrow 0$. In fact, $\alpha$ is
given by the balance between the competing effects of
tunnelling ($\propto \delta_JN$), which enforces a 
well-defined inter-trap phase, and inverse 
compressibility ($\propto E_c$), which favors the formation
of separate number states in the two traps, 
thus separating the two macroscopic wave functions.

A small finite value of $\Delta U$ in eqn (\ref{Hnormal})
displaces the equilibrium position
of the harmonic oscillator. Noticeably, the ground state is 
a \emph{coherent} state with harmonic evolution of the average phase
in time, 
\begin{equation}
\phi_0(\tau)=-\frac{\Delta U}{\hbar\omega_J}\sin{(\omega_J\, \tau)},
\end{equation}
whereas $\alpha$ is unchanged. 
The Gaussian probability density characteristic of the ground state
for $\Delta U=0$ now is simply carried back and forth in $\phi$
space in the
same motion as the expectation value $\phi_0(\tau)$.

\begin{figure}
\setlength{\unitlength}{1 cm}
\begin{picture}(8.5,6.5)
\put(2.6,0.0){\epsfig{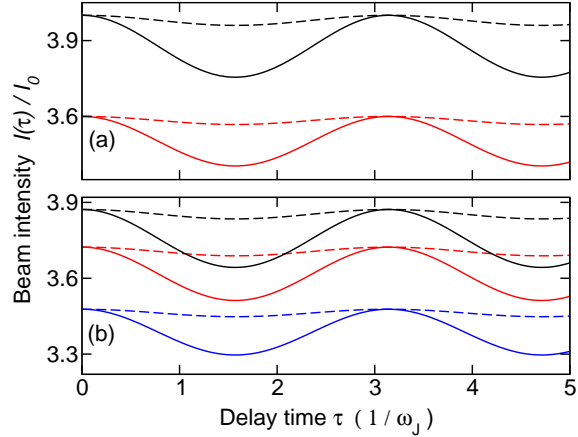}}
\end{picture}
\caption{(Color online).  Beam intensity $I(\tau)/I_0$
vs delay time $\tau$, for $\Delta U/\hbar\omega_J=$ 0.2, 0.5 (dashed 
and solid lines, respectively). (a) $T=$ 0 and $\alpha=$ 1, 0.8
(black and red [light gray] lines, respectively). (b) $\alpha(T=0)=$ 0.94 and 
$k_B T/\hbar\omega_J = $ 0, 1, 2 (black, red [light gray], and blue 
[dark gray] lines, respectively).
\label{fig2}}
\end{figure}

This key feature allows for
directly monitoring $\tau$-dependent 
plasma oscillations of frequency $\omega_J$
through the photon correlation measurement
(cf.~Fig.~\ref{fig2}).
We evaluate the effect of thermal fluctuations on 
$\alpha$ via the formula
\begin{equation}
\alpha(T) = \frac{ \sum_n \alpha_n 
\exp{[-\beta E_n]} } { \sum_n \exp{[-\beta E_n]} },
\end{equation}
where $\beta = 1/k_{\text{B}} T$ and 
$2(\alpha_n-1)=\left<(\Delta\phi)^2\right>_n$ is the
variance of $\phi$ in 
the $n$th excited state whose energy is $E_n$. At low $T$, 
the excited states may be approximated as those
of the harmonic oscillator, giving 
\begin{equation}
\alpha(T)=1-\sqrt{\frac{E_c}{2\delta_J N}}
\left(\frac{1}{2}+\frac{1}{e^{\beta \hbar \omega_J}
-1}\right).
\end{equation}
The above results are summarized by the formula
\begin{equation}
I(\tau)=2I_0\!\left[1+\alpha(T)\cos{\!\left(\frac{\Delta U}{\hbar
\omega_J}\sin{\omega_J\,\tau}\right)} \right],
\label{interfe3}
\end{equation}
which is valid for $E_c/\delta_J N \ll 1$. 

For small dipole energy variations, $\Delta U/ \hbar \omega_J \ll 1$,
the oscillating part within the square brackets of eqn (\ref{interfe3})
may be written as 
$-\alpha(T)/2(\Delta U/\hbar \omega_J)^2\sin^2{\omega_J\,\tau}$.
This shows that the visibility $\alpha(T)$ of fringes, which oscillate
like $\sin^2{\omega_J\,\tau}$, is modulated by
the experimentally tunable factor $(\Delta U/\hbar \omega_J)^2/2$.
The dependence of $I(\tau)$ on $\Delta U$ 
is illustrated in Fig.~\ref{fig2} for two values of
$\Delta U/\hbar\omega_J$. As $\Delta U/\hbar\omega_J$ is increased
[from 0.2 (dashed lines) to 0.5 (solid lines)], the amplitude of
oscillations of $I(\tau)$ shows a strong non-linear enhancement, 
providing a clear signature of Josephson oscillations.  
The oscillation amplitudes are 
larger for higher values of $\alpha$ [cf.~Fig.~\ref{fig2}(a)], 
and fairly robust against thermal smearing 
[cf.~Fig.~\ref{fig2}(b)]. In fact, Fig.~\ref{fig2}(b) shows
that the oscillation of $I(\tau)$ is still clearly resolved 
for temperatures as high as
$T\approx \hbar\omega_J/k_B$. At even higher temperatures
$\alpha(T)$ displays anharmonic effects \cite{Pitaevskii2001}, with
$\alpha(T)\rightarrow 0$ as $T\rightarrow \infty$.

The AC Josephson effect cannot be observed within our scheme.
In fact, for large values of $\Delta U$,
the term proportional to $\cos{\phi}$ appearing in the
Hamiltonian (\ref{H}) may be neglected in first approximation,
and the ground state wave function is a plane wave, $(2\pi)^{-1/2}
\exp{[i\bar{n}\phi]}$, where $\bar{n}$ is the integer closest to
$-\Delta U /E_c$. Since the probability density, $(2\pi)^{-1}$, is constant,
the phase is distributed randomly and
the visibility is zero. Therefore, the correction to $\alpha$ 
coming from the inclusion 
in the calculation of the term neglected 
in (\ref{H}) will be small and fragile 
against fluctuations.

\section{Conclusions}

The venerated topic of exciton Bose Einstein condensation
is facing a rebirth as its investigation
is fueled by advances in novel materials and technologies,
as for double-layer semiconductors and graphene. 
Indeed, growing evidence shows that the concept of exciton condensation
is a paradigm of many-body behavior.
The observation of the coherence properties
of the condensate, including superfluidity, 
is an important long-term goal of this field.
In this Chapter we have explained how the exciton analogues of 
Andreev reflection and Josephson oscillations
may be linked to measurable quantities.
We hope that these ideas may stimulate further experiments
along this path.

\appendix 

\bibliographystyle{OUPnum}
%\bibliography{revised}

\end{document}